\documentclass[aps,pre,reprint,longbibliography,showpacs,showkeys,amsmath,amssymb,twocolumn,nofootinbib]{revtex4}
\usepackage{graphicx}
\usepackage{amsfonts}
\usepackage{url}
\usepackage{epstopdf}
\usepackage{color}
\usepackage{bm}
\usepackage[colorlinks=true,linkcolor=blue,citecolor=red]{hyperref}%
\usepackage{bm}% bold math
\usepackage{amssymb}
\usepackage{amsmath}
\usepackage{ulem}
\usepackage{comment} % Include hidden comments

\usepackage{color,soul}  %%%%%%%%%%%%%%%%%%%%%
\setstcolor{red}         %%%%%%%%%%%%%%%%%%%%%
\newcommand{\be}{\begin{equation}}
\newcommand{\ee}{\end{equation}}
\newcommand{\ba}{\begin{eqnarray}}
\newcommand{\ea}{\end{eqnarray}}
\newcommand{\besu}{\begin{subequations}}
\newcommand{\esu}{\end{subequations}}

\DeclareMathOperator{\erfc}{erfc}

\def\atan{{\rm atan}}

\def\pa{\partial\Omega}

\def\P{{\mathbb P}}
\def\R{{\mathbb R}}

\def\T{{\mathcal T}}

\def\erf{\mathrm{erf}}
\def\erfc{\mathrm{erfc}}

\begin{document}
\title{First-encounter time of two diffusing particles in confinement}
\author{F. Le Vot}
\affiliation{
Departamento de F\'{\i}sica and Instituto de Computaci\'on Cient\'{\i}fica Avanzada (ICCAEx) \\
Universidad de Extremadura, E-06071 Badajoz, Spain}

\author{S.~B. Yuste}
\affiliation{
Departamento de F\'{\i}sica and Instituto de Computaci\'on Cient\'{\i}fica Avanzada (ICCAEx) \\
Universidad de Extremadura, E-06071 Badajoz, Spain}

\author{E. Abad}
\affiliation{
Departamento de F\'{\i}sica Aplicada and Instituto de Computaci\'on Cient\'{\i}fica Avanzada (ICCAEx) \\
Centro Universitario de M\'erida \\ Universidad de Extremadura, E-06800 M\'erida, Spain}

\author{D.~S. Grebenkov}
\affiliation{
Laboratoire de Physique de la Mati\`{e}re Condens\'{e}e (UMR 7643), \\
CNRS -- Ecole Polytechnique, IP Paris, 91128 Palaiseau, France;}
\affiliation{Institute of Physics \& Astronomy, University of Potsdam,
14476 Potsdam-Golm, Germany}

\begin{abstract}
We investigate how confinement may drastically change both the
probability density of the first-encounter time and the related
survival probability in the case of two diffusing particles.  To
obtain analytical insights into this problem, we focus on two
one-dimensional settings: a half-line and an interval.  We first
consider the case with equal particle diffusivities, for which exact
results can be obtained for the survival probability and the
associated first-encounter time density over the full time domain.  We
also evaluate the moments of the first-encounter time when they exist.
We then turn to the case when the diffusivities are not equal, and
focus on the long-time behavior of the survival probability.  Our
results highlight the great impact of boundary effects in
diffusion-controlled kinetics even for simple one-dimensional
settings, as well as the difficulty of obtaining analytic results as
soon as translational invariance of such systems is broken.
\end{abstract}


%02.50.-r       (Probability theory, stochastic processes, and statistics)
%05.40.-a 	Fluctuation phenomena, random processes, noise, and Brownian motion
%02.70.Rr       (General statistical methods)
%05.10.Gg 	Stochastic analysis methods (Fokker-Planck, Langevin, etc.)

%02.50.Ey 	Stochastic processes  (Probability theory, stochastic processes, and statistics)


\maketitle

\section{Introduction}

As most chemical reactions are encounter-controlled, the
first-encounter time (FET) of the reactants is one of the central
quantities characterizing diffusion-influenced reactions.  The first
study of the FET goes back to Smoluchowski, who reduced a many-body
reaction problem of two species (i.e., bimolecular reactions) with a
vast excess of one species, to the problem of two diffusing reactive
particles \cite{Smoluchowski1917}.  By selecting a coordinate system
that follows one of the diffusing particles, the original problem is
reduced to the simpler problem of a single particle diffusing towards
a static target (or sink).  Smoluchowski solved this problem and
determined the survival probability, whence the probability density of
the first-passage time to the target (here equivalent to the FET),
from which the associated reaction rate immediately follows.

Since Smoluchowski's seminal work, first-passage times to fixed
targets have been thoroughly investigated for various kinds of
diffusion processes, chemical kinetics, and geometric settings
\cite{Rice,Lauffenburger,Redner,Schuss,Metzler,Oshanin,Sano79,Agmon90,Levitz06,Condamin07,Grebenkov07,Benichou10,Benichou10b,Grebenkov10a,Grebenkov10b,Benichou11,Bressloff13,Benichou14,Galanti16,Guerin16,Lanoiselee18,Grebenkov19,Grebenkov19d}.
In particular, when the fixed target is small, one deals with the
so-called narrow escape problem, for which many asymptotic results
have been derived
\cite{Holcman04,Schuss07,Benichou08,Pillay10,Cheviakov10,Cheviakov12,Rupprecht15,Grebenkov17,Grebenkov18b}
(see also a review \cite{Holcman14}).  Another well-explored research
direction concerns multiple particles diffusing on infinite lattices
or in Euclidean spaces.  This general setting allows one to
investigate elaborate chemical reactions involving various species,
the effect of inter-particle interactions (e.g., excluded volume), and
cooperativity effects when, for instance, several predators hunt for a
prey
\cite{Szabo1988,Redner1999,Blythe2003,Yuste08,Borrego09,Oshanin2009,Majumdar10,LeDoussal19}.
In this context, one clearly identifies two types of problems, i)
those where any pair of particles can interact with each other as long
as such interactions are not precluded by geometric constraints, and
ii) those where particles of a given species (usually the majority
species) do not interact with one another, but do so with a target
particle or with a set of targets. The first type is well exemplified
by binary reactions such as one-species and two-species
coalescence/annihilation reactions \cite{Privman}, whereas the second
type includes the so-called target problem and the trapping problem,
as well as variants thereof
\cite{Blythe2003,Bray13,Kayser83,Kayser84,Torquato86,Lee89,Torquato91,Torquato97,Kansal02,Yuste01,Acedo02,Yuste04,Yuste05,Yuste06,Yuste07,Yuste08,Borrego09,Abad12,Abad13}.
In particular, the generic question on how the mobility of a target or
a trap impacts the reaction rate has long been a subject of interest
\cite{Blythe2003,Yuste08,Borrego09,Moreau03,Moreau04,Bray13,Bramson88,Bramson91,Bray02,Yuste05}.
The third direction regroups numerical works, in which
diffusion-reaction processes are modeled by molecular dynamics or
Monte Carlo simulations \cite{McGuffee10,Ghost16,Samanta16}.  While
such approaches are admittedly the most realistic ones, they often
lack analytic insights which are often of great help for the intuitive
understanding and systematic characterization of diffusion-reaction
processes.

Quite surprisingly, the influence of confinement onto the distribution
of the FET between two diffusing particles and the consequent chemical
reactions is much less studied.  The evident consequence of the
presence of a confining boundary is the translational symmetry
breaking that prohibits the reduction of two diffusing particles to a
single particle diffusing towards a static target.  One therefore has
to describe the dynamics of two particles inside a confining domain,
and the solution of the relevant diffusion-reaction equations becomes
much more sophisticated.  We are aware of only few works dealing with
such problems, and they are concerned with the simplest possible
scenario of infinite reaction rate: the reaction takes place with unit
probability upon encounter. In situations when the consequent fate of
the products of the reaction are not of primary interest (e.g., if the
reaction products are inert), one may formally consider that any two
diffusing walkers annihilate irreversibly upon encounter but do not
interact otherwise.  Fisher coined the term ``vicious walkers'' for
such non-intersecting walks \cite{Fisher,Lawler}. Upon Fisher's
systematic study of their statistical properties, vicious walkers
became an important paradigm in statistical physics. Fisher's original
formulation was in terms of lattice walks, but the diffusive limit of
the latter is often considered, as it greatly simplifies the
mathematical treatment. In this diffusive approximation, Bray and
Winkler studied one-dimensional vicious walkers in a potential,
including the case of an interval \cite{Bray04}. Further references on
vicious walkers in finite systems are given at the end of
Sec. \ref{sec:equations}.

As far as other works are concerned, Amitai {\it et al.} estimated the
mean FET between two ends of a polymer chain by computing the mean
time for a Brownian particle to reach a narrow domain in the polymer
configuration space \cite{Amitai2012}.  Tejedor {\it et al.}
investigated diffusion of two particles with equal diffusivities on an
interval with either absorbing or reflecting boundary conditions and
computed two quantities: the probability that the random walkers meet
before one of them is removed at the absorbing interval boundaries,
and the typical encounter time of the two walkers in the presence of
reflecting boundaries \cite{Tejedor11}.  The related epidemic
spreading problem has been discussed in \cite{Giuggioli13}.  Tzou {\it
et al.} studied the mean FET for two particles diffusing on a
one-dimensional interval by solving numerically the underlying
diffusion equations \cite{Tzou2014}.  In particular, they studied
whether a mobile trap can improve capture times over a fixed trap.
Even for such a simple geometric setting, an analytical solution of
the problem was not provided.  More recently, Lawley and Miles
computed the mean FET for a very general diffusion model with many
{\it small} targets that can diffuse either inside a three-dimensional
domain, or on its two-dimensional boundary, their diffusivities are
subject to random fluctuations, while their reactivity can be
stochastically gated \cite{Lawley19}.  However, a systematic study of
the first-encounter time {\it distribution} for diffusing particles in
confinement is still missing.

In this paper, we consider two Brownian particles $A$ and $B$
diffusing inside a bounded domain with reflecting boundary, and
investigate the probability of the particles not having met each other
up to a given time $t$.  This quantity is the survival probability for
a pair of two molecules with infinite reactivity so that their first
collision leads to a chemical reaction: $A + B \rightarrow C$.  In
chemical kinetics, the survival probability can be interpreted as the
fraction of particles still reactive at time $t$ with respect to the
initial number of particles.  The survival probability determines
other important quantities: the probability density of the FET, its
mean value and higher-order moments, as well as the reaction rate.

For two diffusing spherical particles without confinement (i.e., in
$\R^d$), the survival probability and related quantities are functions
only of the initial distance between the centers of the particles, of
the sum of their radii, and of the sum of their diffusion constants.
In contrast, confinement induces new length scales involving distances
between the particles and the reflecting boundary, and changes
chemical kinetics, particularly at long times at which the typical
distance traveled by particles is comparable with the system size.
Even though most chemical reactions occur under confinement, its
impact on the survival probability and related quantities remains
poorly understood.  In view of these shortcomings, our work aims to
shed further light on the role of confinement in bimolecular
diffusion-limited reactions.  To this end, we will use both analytical
tools and numerical simulations. In contrast with some previous works,
our analysis will extend beyond the long-time asymptotic regime
whenever possible, since the influence of the domain boundaries may
already become apparent for comparatively short times. The interaction
with the reflecting boundaries will not only alter the value of the
mean FET, but also affect higher order moments, which assess the
impact of trajectory-to-trajectory fluctuations and the statistical
significance of the mean FET.

The paper is organized as follows.  In Sec.~\ref{sec:equations}, we
formulate diffusion-reaction problem and summarize the main known
theoretical results that are relevant for our study.  In
Sec.~\ref{sec:half-line}, we consider the most studied case of two
particles diffusing on a half-line $\R_+$ with reflecting endpoint at
$0$.  We provide the exact solutions for the survival probability, the
FET probability density, as well as the moments of the FET.  While
some aspects of this first-encounter problem have been analyzed in
earlier works, to our knowledge many of the reported properties are
new. In Sec.~\ref{sec:interval}, we explore the FET problem for
diffusion on an interval $(0,L)$ with reflecting endpoints.  In spite
of the apparent simplicity of this geometric setting, much fewer
analytical results are known, especially for unequal diffusivities.
First, we consider in Sec.~\ref{sec:int_equal} the problem with equal
diffusivities, for which an exact solution for the survival
probability, the FET density and moments can be obtained in the form
of spectral expansions.  Next, we discuss in
Sec. \ref{sec:int_different} the case of unequal diffusivities; even
though the exact solution is unknown, we investigate its long-time
decay by studying the behavior of the smallest eigenvalue of the
Laplace operator in this setting.  In particular, we show that the
associated decay time depends on both $D_1$ and $D_2$ in a complex
way.  Finally, our main conclusions are summarized in
Sec.~\ref{Sec:Conclusions}.

\section{Summary of known results}
\label{sec:equations}

In this section, we summarize some theoretical results on the
first-encounter time in the one-dimensional case.  Even though these
results are known, they are dispersed in the literature and not easily
accessible.  The problem of the first-encounter time of two diffusing
particles in the one-dimensional case is very specific and different
from higher-dimensional settings: (i) the particles can be point-like
and still meet with probability one; (ii) the particle cannot overpass
each other without meeting, i.e., their initial order is always
preserved.  These two properties allow one to derive some analytical
solutions which are not available in higher dimensions.  Nevertheless,
the one-dimensional setting provides a solid theoretical background
and some intuition on the FET and related properties in higher
dimensions.  Beyond this immediate justification, the considered
setting is also directly related to several fundamental problems in
statistical physics related to fibrous structures \cite{DeGennes},
polymer networks \cite{EssamGuttmann}, wetting transitions,
dislocations, and melting \cite{Fisher,HuseFisher}.

We consider the problem of the first-encounter time for two
independent point-like particles diffusing with diffusion coefficients
$D_1$ and $D_2$ on a domain $\Omega \subset \R$ with reflections on
its boundary $\pa$.  Let us respectively denote by $x_1$ and $x_2$ the
starting positions of the particles and assume that $x_1 \geq x_2$
without loss of generality.  The first-encounter time $\T$ of these
particles is a random variable characterized by a cumulative
probability distribution, $\P\{ \T < t\}$, or, equivalently, by the
survival probability $S(t|x_1,x_2) = \P\{ \T > t\}$.  As the encounter
depends on the positions of both particles, it is natural to consider
their joint dynamics in the phase space $\Omega \times \Omega$.  In
particular, the survival probability satisfies the backward diffusion
equation:
\begin{equation}  \label{eq:Sgeneral_def}
\frac{\partial S}{\partial t} = \left(D_1 \frac{\partial^2}{\partial x_1^2} + D_2 \frac{\partial^2}{\partial x_2^2}\right) S
\qquad (x_1,x_2)\in \Omega \times \Omega,
\end{equation}
subject to the initial condition $S(t=0|x_1,x_2) = 1$ for $x_1
\ne x_2$, the Neumann boundary condition on the reflecting boundary
$\pa$ (there is no net diffusive flux across the boundary), and the
Dirichlet boundary condition $S(t|x_1,x_2=x_1) =0$, meaning an
immediate reaction upon the first encounter.  Once the survival
probability is found, one easily gets the FET probability density
\begin{equation}  \label{eq:H_def}
H(t|x_1,x_2) = - \frac{\partial}{\partial t} S(t|x_1,x_2)
\end{equation}
and the associated moments:
\begin{equation}   \label{eq:Tkmean}
\langle \T^k \rangle = k \int\limits_0^\infty dt\, t^{k-1} \, S(t|x_1,x_2)
\end{equation}
(note that, depending on the problem at hand, such integrals may not
converge, i.e., some moments or even all of them can be infinite).
Alternatively, integrating the Eq.~\eqref{eq:Sgeneral_def} over time
from $0$ to $\infty$, assuming that $S(t)\to 0$ for $t\to\infty$, and
taking into account Eq.~\eqref{eq:Tkmean} for $k=1$, one finds that
the mean FET $\langle \T \rangle$ satisfies
\begin{equation}  \label{eq:Tgeneral_def}
-1 = \left(D_1 \frac{\partial^2}{\partial x_1^2} + D_2 \frac{\partial^2}{\partial x_2^2}\right) \langle \T \rangle
\qquad (x_1,x_2)\in \Omega \times \Omega,
\end{equation}
the Neumann boundary condition $\partial \langle \T \rangle/\partial
n=0$ on the reflecting boundary $\pa$, and the annihilation reaction
condition $ \langle \T \rangle(x_1,x_2)=0$ when $x_1=x_2$.  Similar
equations are available for higher-order integer moments.

It is natural to assume that the domain $\Omega$ is connected
(otherwise the particles could not move from one component to another,
and the problem would be trivially reduced to that in a single
component).  As a consequence, there are only three possible settings:
(i) $\Omega = \R$, (ii) $\Omega$ is a half-line, and (iii) $\Omega$ is
a finite interval.  The first case, also known as ``the diffusing
cliff'' in the literature on first-passage processes \cite{Redner}, is
well known; as already anticipated, Smoluchowski's argument reduces
the original problem to that of a single effective particle diffusing
with diffusivity $D_1 + D_2$ towards a fixed target.  The related
survival probability is retrieved by solving the simple diffusion
equation on a half-line:
\begin{equation}
\label{freesurvival}
S_{\rm free}(t|x_1,x_2) = \erf\biggl(\frac{\delta}{\sqrt{4(D_1+D_2)t}}\biggr),
\end{equation}
where $\erf(z)$ is the error function and $\delta \equiv x_1-x_2$ is
the initial separation distance between the particles.  Thus, the
statistics of FET depends only on the sum of diffusion coefficients
and on the initial inter-particle distance.  In particular, the
long-time decay of Eq. \eqref{freesurvival} is obtained from the
asymptotic behavior of the error function:
\begin{equation}
\label{longtimesurvival}
S_{\rm free}(t|x_1,x_2) \sim \frac{\delta}{\sqrt{\pi (D_1+D_2) t}} \qquad t\gg \frac{\delta^2}{D_1+D_2} \,.
\end{equation}
The FET probability density follows upon deriving
Eq. \eqref{freesurvival} with respect to time:
\begin{equation}
\label{freedensity}
H_{\rm free}(t|x_1,x_2) = \frac{\delta \, \exp\bigl(-\frac{\delta^2}{4(D_1 + D_2)t}\bigr)}{\sqrt{4\pi (D_1+D_2) t^3}}  \,,
\end{equation}
implying the long time behavior
\begin{equation}
\label{asyfreedensity}
H_{\rm free}(t|x_1,x_2) = \frac{\delta \, t^{-3/2}}{\sqrt{4\pi (D_1+D_2)}}  \qquad  t\gg \frac{\delta^2}{D_1+D_2} \,.
\end{equation}
The mean FET, as well as higher-order moments, are infinite.  In fact,
even though the particles meet with probability $1$, long trajectories
before encounter provide dominant contributions to these moments.  In
the following, these results for the infinite system will be used as a
reference for the half-line and for the finite interval.

The solutions for both a half-line ($\Omega = \R_+$) and an interval
($\Omega = (0,L)$) are obtained by stretching one of the coordinates
in order to reduce Eq. (\ref{eq:Sgeneral_def}) to a diffusion equation
with equal diffusivities on a planar domain (see below).  The
half-line case has been extensively studied by Redner {\it et al.}
(see \cite{Redner,Redner1999} and references therein).  In particular,
the long-time asymptotic behavior of the survival probability was
provided in \cite{Redner}.  Despite such extensive studies, the main
focus so far was clearly on the long-time asymptotics, to the extent
that we have not been able to find direct exact solution of this
problem, but rather the solution of an equivalent problem with a
single particle diffusing in a wedge with absorbing boundaries.  For
this reason, not only do we provide this solution in
Sec. \ref{sec:half-line}, but we also analyze some interesting
features characterizing the moments of the FET and the transient
behavior of the survival probability.

Apart from the above results, problems of vicious walkers under
geometric constraints remain widely unexplored. In Ref. \cite{Fisher},
Fisher considered the effect of an absorbing wall on the reunion
statistics of identical or dissimilar walkers; the reunion of
dissimilar walkers was subsequently studied by Fisher and Gelfand
\cite{FisherGelfand}.  Forrester \cite{Forrester} investigated finite
size effects introduced by periodic boundaries.

The problem of diffusion on an interval has been studied much less
extensively.  Bray and Winkler considered the problem of $N$ vicious
walkers in different settings, including an interval with reflecting
endpoints \cite{Bray04}, thereby generalizing previous results by
Krattenthaler \cite{Krattenthaler}.  However, they restricted their
analysis to the case of identical particles and focused on the
asymptotic long-time behavior of the survival probability.
More recently, Forrester {\it et al.} \cite{ForrMajuSchehr} considered
from another viewpoint the reunion statistics of non-intersecting
Brownian motions (i.e., surviving vicious walks) on an interval with
periodic, reflecting and absorbing boundary conditions. In the
absorbing case, they showed that the normalized reunion probability is
related to the statistics of the outermost Brownian path on the half
line. These results were further explored by Liechty \cite{Liechty}.
We will revisit the problem of diffusion on an interval in
Sec. \ref{sec:interval}.

\section{Half-line}
\label{sec:half-line}

\subsection{Survival probability}

We consider the problem of the first-encounter time for two particles
started from points $x_1 > x_2$ and diffusing with diffusion
coefficients $D_1$ and $D_2$ on the positive half-line $\R_+$ with
reflections at $0$.  As mentioned in Sec. \ref{sec:equations}, this
problem is equivalent to two-dimensional diffusion in the
half-quadrant (or wedge of angle $\pi/4$) with reflecting horizontal
axis and the absorbing diagonal.  Rescaling the coordinate of the
second particle by $\sqrt{D_1/D_2}$, i.e., setting new coordinates
$y_1 = x_1$ and $y_2 = x_2 \sqrt{D_1/D_2}$, one maps the original
problem onto the problem of isotropic diffusion (with diffusion
coefficient $D_1$) in a wedge $\Omega_0 = \{ 0< r < \infty,~ 0 <
\theta < \Theta\}$, with the reflecting ray at $\theta = 0$ and
absorbing ray at $\theta = \Theta$, with the wedge angle
\begin{equation}  \label{eq:Theta}
\Theta = \atan(\sqrt{D_1/D_2}).
\end{equation}
The latter ray accounts for the encounter condition $y_1 = x_1 = x_2 =
y_2 \sqrt{D_2/D_1}$ when two particles meet.  The initial position in
the wedge is determined by polar coordinates $(r_0,\theta_0)$ with
$r_0 = \sqrt{y_1^2 + y_2^2} = \sqrt{x_1^2 + x_2^2 D_1/D_2}$ and
$\theta_0 = \atan(y_2/y_1) = \atan(x_2\sqrt{D_1/D_2}/x_1)$.  Note that
the assumed condition $x_1 > x_2$ implies $\theta_0 < \Theta$.

In polar coordinates, the survival probability reads $S(t|x_1,x_2) =
U(t|r_0,\theta_0)$, where the function $U(t|r_0,\theta_0)$ satisfies
the diffusion equation,
\begin{equation}
\frac{\partial}{\partial t} U = D_1 \Delta U ,
\end{equation}
subject to two boundary conditions:
\begin{align}
\biggl(\frac{\partial U}{\partial \theta} \biggr)_{\theta = 0} &= 0,  \qquad
U_{\theta = \Theta} = 0 .
\end{align}
Due to the symmetry, one can replace $\Omega_0$ by a twice larger
wedge $\Omega = \{ 0< r < \infty,~ -\Theta < \theta < \Theta\}$, with
Dirichlet condition $U=0$ on its boundary.

The radial Green's function for a wedge domain $\Omega$ was provided in
\cite{Carslaw} (see p. 379)
\begin{align}  \nonumber
G(r,\theta,t & |r_0,\theta_0) = \sum\limits_{n=1}^\infty \frac{e^{-(r^2+r_0^2)/(4D_1t)}}{D_1t}  I_{\nu_n}(rr_0/(2D_1t)) \\
& \times \frac{1}{2\Theta} \sin(\nu_n(\theta + \Theta)) \sin(\nu_n(\theta_0 + \Theta)) ,
\end{align}
where $\nu_n = \pi n/(2\Theta)$, and $I_\nu(\cdot)$ is the modified
Bessel function of the first kind.  In order to use the result of
Ref. \cite{Carslaw}, here we have considered the wedge of angle
$2\Theta$ and we have then shifted the angular coordinate by $\Theta$.
The integral of this formula over the arrival point $(r,\theta)$
yields the survival probability
\begin{align}  \nonumber
& S(t|r_0,\theta_0) = \int\limits_{-\Theta}^{\Theta} d\theta \int\limits_0^\infty dr\, r \, G(r,\theta,t|r_0,\theta_0) \\
\label{sp-HL}
& = 4 \sum\limits_{n=1}^\infty \frac{1-(-1)^n}{\pi n} \sin(\nu_n(\theta_0 + \Theta)) \, R_{\nu_n}\bigl(r_0/\sqrt{D_1t}\bigr) ,
\end{align}
where
\begin{align}  \nonumber
R_\nu(z) &= e^{-z^2/4} \int\limits_0^\infty dx\, x \, e^{-x^2} I_{\nu}(xz) \\
& = \frac{\sqrt{\pi}}{8} z e^{-z^2/8} \biggl(I_{\frac{\nu-1}{2}}(z^2/8) + I_{\frac{\nu+1}{2}}(z^2/8)\biggr).
\end{align}
Note that this function approaches $1/2$ in the limit $z\to \infty$
and behaves as $R_\nu(z) \propto z^\nu$ as $z\to 0$.  The probability
density of the FET then reads
\begin{align}  \nonumber
H(t|r_0,\theta_0) & = \frac{2r_0}{\sqrt{D_1}\, t^{3/2}}
\sum\limits_{n=1}^\infty \frac{1-(-1)^n}{\pi n} \sin(\nu_n(\theta_0 + \Theta)) \\
\label{fpd-HL}
& \times  R'_{\nu_n}\bigl(r_0/\sqrt{D_1t}\bigr) ,
\end{align}
where
\begin{equation}
R'_\nu(z) = \frac{\sqrt{\pi}\, \nu}{8} e^{-z^2/8} \biggl( I_{\frac{\nu-1}{2}}(z^2/8) - I_{\frac{\nu+1}{2}}(z^2/8)\biggr).
\end{equation}
Note that if one sets $x_1=x_2+\delta$ in Eqs. \eqref{sp-HL} and
\eqref{fpd-HL} and takes the limit $x_2\to\infty$, one recovers
Eqs. \eqref{freesurvival} and \eqref{freedensity} for an infinite
system.

As discussed by Redner \cite{Redner}, the survival probability decays
asymptotically as
\begin{equation}  \label{eq:S_Redner}
S \propto (r_0/\sqrt{D_1t})^{\pi/(2\Theta)} \propto t^{-\pi/(4\Theta)} \qquad (t\to \infty) ,
\end{equation}
so that the probability density decays as $H \propto
t^{-\pi/(4\Theta)-1}$.  Note that, since one always has
$\pi/(4\Theta)>1/2$, the long time decay of both $S$ and $H$ is faster
than in the free case [cf.  Eqs. \eqref{longtimesurvival} and
\eqref{asyfreedensity}]. This simply reflects the fact that the
boundary favors a faster reaction.

Surprisingly, the early time behavior of the solution seems to have
received much less attention in the literature in comparison with the
long time asymptotics.  In fact, at short time, subleading corrections
characterize the departure of Eq. \eqref{sp-HL} from the free solution
\eqref{freesurvival} arising from the perturbation introduced by the
reflecting boundary. In a recent work addressing the diffusion of a
random walker in a two-dimensional wedge with absorbing boundaries
\cite{Chupeau15}, Chupeau {\it et al.} noted that the series
representation \eqref{sp-HL} does not allow one to extract the
short-time behavior because the use of the asymptotic expansion of the
modified Bessel functions would yield a divergent series. Instead,
Chupeau {\it et al.} suggested to resort to the integral
representation of the modified Bessel functions to derive an
alternative form of $S(t|r_0,\theta_0)$. This yields an exact
expression in terms of complementary error functions and an integral
remainder, but for some specific values of $\Theta$ the latter
disappears \cite{Dy2008}.  For example, when $\Theta=\pi/4$ (which
corresponds to $D_1=D_2$), one finds
\begin{equation}
S(t|r_0,\theta_0,\Theta=\pi/4) = \erf\bigl(\sqrt{2y}\sin\varphi_0\bigr) \, \erf\bigl(\sqrt{2y}\cos\varphi_0\bigr),
\end{equation}
where $\varphi_0=\theta_0+\Theta$ and $y=r_0^2/(8 D_1 t)$.  In terms
of $x_1$ and $x_2$, this gives
\begin{subequations}
\begin{eqnarray}
\label{Seq}
&& S(t|x_1,x_2,D_2=D_1)= \nonumber \\
&&= \erf\left(\frac{x_1-x_2}{\sqrt{8D_1 t}}\right) \erf\left(\frac{x_1+x_2}{\sqrt{8D_1 t}}\right)= \label{prodform} \\
&&= S_{\rm free}(t|x_1,x_2,D_2=D_1) \, \erf\left(\frac{x_1+x_2}{\sqrt{8D_1 t}}\right)= \nonumber \\
&&= S_{\rm free}(t|x_1,x_2,D_2=D_1)- \nonumber  \\
&&-S_{\rm free}(t|x_1,x_2,D_2=D_1) \, \erfc\left(\frac{x_1+x_2}{\sqrt{8D_1 t}}\right). \label{separate-cont}
\end{eqnarray}
\end{subequations}
The last term in Eq. \eqref{separate-cont} quantifies the effect of
the boundary on the survival probability.  

Another case, in which a closed-form expression in terms of error
functions is available is $\Theta=\pi/6$, corresponding to $D_2=3D_1$:
\begin{eqnarray}
&& S(t|x_1,x_2,D_2=3D_1) = \erf\left(\frac{x_1-x_2}{\sqrt{16D_1 t}}\right)+\nonumber \\
&& + \erf\left(\frac{x_1+x_2}{\sqrt{16D_1 t}}\right) - \erf\left(\frac{x_1}{\sqrt{4D_1 t}}\right),
\label{separate-cont2}
\end{eqnarray}
where the first term on the rhs corresponds yet again to the free
solution with diffusion coefficient $D_1 + D_2 = 4D_1$.  Equations
\eqref{Seq} and \eqref{separate-cont2}, as well as the alternative
general representation of the survival probability obtained in
\cite{Chupeau15}, provide a good starting point to study the role of
the boundary by comparing the early-time behavior of the solution with
that of Eq. \eqref{freesurvival}, as discussed in 
Sec. \ref{subsec:free-case}.

\subsection{Comparison with the free case}
\label{subsec:free-case}

It is instructive to compare the behavior of $S$ and $H$ with their
counterparts for the free case.  Clearly, $\Delta S\equiv S_{\rm
free}(t|x_1,x_2)-S(t|x_1,x_2)\ge 0$ at all times.  As time goes by,
$\Delta S$ first increases from its initial value $\Delta S(t=0) = 0$,
then attains a maximum value at a time $t_{\star}$, and finally
decreases until it eventually vanishes in the limit $t\to\infty$.  One
has $\left. \partial \Delta S/\partial t \right|_{t=t_{\star}} = 0$,
implying $H(t_{\star}|x_1,x_2) = H_{\rm free}(t_{\star}|x_1,x_2)$.
Thus, the value of $t_{\star}$ can be obtained by solving this
equation numerically.

\begin{figure}
\centering
\includegraphics[width=88mm]{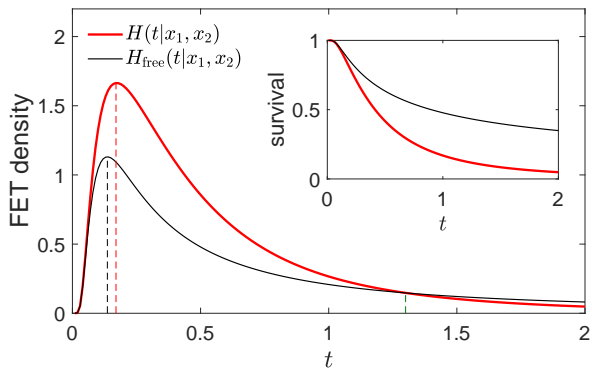} 
\caption{
FET density for the free case (thin black line) and for the half-line
(thick red line).  The chosen parameter values are $x_2=0.5$,
$\delta=3$, $D_1=1$, and $D_2=10$.  Vertical dashed lines indicate the
times $t_{\rm free}$ and $t_{\rm HL}$ when peaks of the FET density
are attained in both cases, as well as the crossing time $t_{\star}$.}
\label{fig:FETdensity}
% [H] = A_Felipe_H_1D_fig();
\end{figure}

Figure \ref{fig:FETdensity} illustrates the typical behavior of the
FET density.  In an infinite system, the FET density reaches a maximum
at $t_{\rm free}=\delta^2/(6(D_1+D_2))$.  In the half-line system, the
density peaks at a later time $t_{\rm HL}$.  Finally, both curves
cross at $t_{\star}$.

When the second particle starts very close to the boundary or exactly
at it ($x_2=0$), one may even have $t_{\rm HL} < t_{\rm free}$ for a
suitable parameter choice, but the difference between these times is
in general small.
Finally, we note that, for $x_2>0$ and a fixed $D_2$, one has
$S(t|x_1,x_2)\to S_{\rm free}(t|x_1,x_2)$ as $D_1\to \infty$; however,
this is not the case if one fixes $D_1$ and then takes the limit
$D_2\to \infty$.

In some special cases, the obtained analytic expressions are more
transparent and therefore easier to interpret.  For instance, for
$D_1=D_2$ and $x_2=0$, Eq. \eqref{prodform} becomes
\begin{eqnarray}
&& S(t|x_1=\delta, x_2=0, D_2=D_1) = \left[\erf\left(\frac{\delta}{\sqrt{8D_1 t}}\right)\right]^2  \nonumber \\
&& = S_{\rm free}^2(t|x_1=\delta, x_2=0, D_2=D_1).
\end{eqnarray}
Thus, the time $t_\zeta$ after which the survival probability is just
a fraction $0<\zeta<1$ of the free solution is simply
\begin{equation}
t_\zeta=\frac{\delta^2}{8D_1 \, [\erf^{-1}(\zeta)]^2}  \,.
\end{equation}
From this equation, one immediately obtains $t_\star=t_{1/2}\approx
0.55 \,\delta^2/(8D_1)$, which is roughly 6.6 larger than $t_{\rm
free}=(1/12) \,\delta^2/(8D_1) \gtrsim t_{\rm HL}$.

\begin{figure}
\centering
\includegraphics[width=88mm]{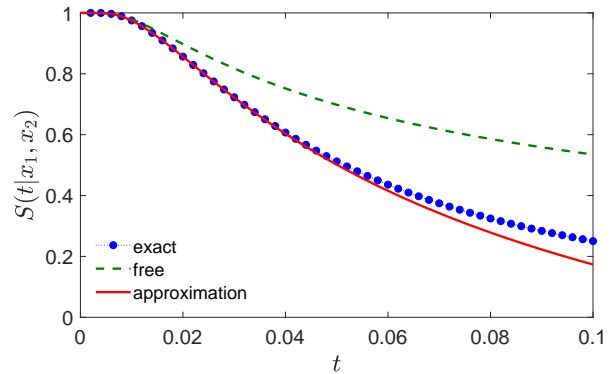} 
\caption{
Time evolution of the survival probability as given by the exact
solution \eqref{sp-HL}, the free solution \eqref{freesurvival} and
Eq. \eqref{earlyapprox}. The chosen parameter values are $x_1=0.9,
x_2=0.1, D_1=1$, and $D_2=5$, leading to $r_0\approx 0.901,
\Theta\approx 0.421$, and $\theta_0\approx 0.05$.}
\label{earlyfigure} 
% A_Felipe_1D_St_fig;
\end{figure}

We close this subsection with a short general discussion on how the
early-time behavior is affected by the boundary. In the case of an
obtuse wedge $\Theta>\pi/4$ (implying $D_1>D_2$), the free solution is
a good approximation up to relatively long times. This holds even if
the particle starts close to the boundary, provided that $\delta$ is
not too small and $D_2$ is not too large. In the case of an acute
wedge $\Theta<\pi/4$ (implying $D_1<D_2$), the time up to which the
free solution is a good approximation, can be significantly
shorter. The free solution is still an acceptable approximation as
long as $\Theta>\pi/6$ but it progressively deteriorates as $D_2$
increases for a fixed $D_1$. If $D_2$ is not too large (such that
$\pi/10<\Theta \le \pi/6$), a better early-time approximation can be
obtained from Eq. (39) in \cite{Chupeau15} by retaining the first two
complementary error functions%
\footnote{
We note that, in the language of Ref. \cite{Chupeau15}, the free
solution corresponds to the asymptotic behavior $ S(y) \simeq 1-
\erfc{\left(\sqrt{2y} \sin\left(2\Theta-\varphi_0\right)\right)}$ of
the survival probability, which holds in the limit $y\equiv
r_0^2/(8D_1 t) \to\infty$ in the parameter range of our problem
($\varphi_0>\Theta$).  This is different from their prediction $ S(y)
\simeq 1- \erfc{\left(\sqrt{2y} \sin\left(\varphi_0\right)\right)}$,
which is not valid in this range. The difference arises because, in
our case, $\psi_+=o(\psi_-)$, rather than $\psi_-=o(\psi_+)$ (note
that $-\psi_+$ and $-\psi_-$ were respectively termed $\psi_1$ and
$\psi_2$ in Ref. \cite{Chupeau15}).}
\begin{equation}
\label{earlyapprox}
S(t|x_1, x_2, D_2, D_1)\approx \underbrace{1-\psi_-}_{S_{\rm free}(t)} - \psi_+ \,,
\end{equation}
with
\begin{equation}
\psi_\pm = \erfc\left(\frac{x_1\pm x_2}{2\sqrt{(D_1+D_2) t}}\right)
\end{equation}
(since the complementary error function is a monotonically decreasing
function, for fixed $x_2$ the correction $\psi_+$ to the free solution
becomes increasingly important with decreasing interparticle distance
$\delta$).  The good accuracy of the approximation
\eqref{earlyapprox} in comparison with the free solution is
illustrated in Fig. \ref{earlyfigure}.

\subsection{Moments of the FET}

We now provide exact expressions for the moments of the FET
density. They can be computed from the survival probability by
applying Eq. (\ref{eq:Tkmean}). Different behaviors are observed
depending on the parameter $\Theta$.  If $\Theta \geq \pi/4$ or,
equivalently, $D_1 \geq D_2$, then one sees from
Eq. \eqref{eq:S_Redner} that the mean FET is still infinite, as in the
free case.  In turn, if $D_1 < D_2$, the mean FET becomes finite and
can be computed from Eq. \eqref{eq:Tkmean} as
\begin{align*}
\langle \T \rangle & = 4 \sum\limits_{n=1}^\infty \frac{1-(-1)^n}{\pi n} \sin(\nu_n(\theta_0 + \Theta)) \hspace*{-1mm}
\int\limits_0^\infty dt \, R_{\nu_n}\biggl(\frac{r_0}{\sqrt{D_1t}}\biggr) ,
\end{align*}
and the last integral reads:
\begin{align*}
& \frac{\sqrt{2\pi}}{32} \, \frac{r_0^2}{D_1} \, \int\limits_0^\infty \frac{dz}{z^{3/2}} e^{-z}
\biggl(I_{\frac{\nu_n-1}{2}}(z) + I_{\frac{\nu_n+1}{2}}(z)\biggr) \\
& = \frac{r_0^2}{2D_1 (\nu_n^2 - 4)} \,.
\end{align*}
The last equality holds for $\nu_n > 2$, which is valid for all $n =
1,2,3,\ldots$ since $\Theta < \pi /4$.  We get thus the mean FET as
\begin{equation}   \label{eq:T_halfline}
\langle \T \rangle = \frac{2r_0^2}{D_1} \sum\limits_{n=1}^\infty \frac{1-(-1)^n}{\pi n (\nu_n^2 - 4)} \sin(\nu_n(\theta_0 + \Theta))   \,.
\end{equation}
Using the identity \cite{Grebenkov20}
\begin{align}  \nonumber
& \sum\limits_{n=1}^\infty \frac{(1-(-1)^n) \sin(\pi n x)}{\pi n (z^2 - \pi^2 n^2)} \\  \label{eq:summation}
& = \frac{\sin(z) - \sin(zx) - \sin(z(1-x))}{2z^2 \sin(z)} \,,
\end{align}
we compute the mean FET as
\begin{equation}   \label{eq:T_halfline2}
\langle \T \rangle = \frac{r_0^2}{4D_1} \biggl(\frac{\cos(2\theta_0)}{\cos(2\Theta)} - 1 \biggr) \,.
\end{equation}
Thanks to trigonometric relations, one can express
\begin{equation}
\cos(2\theta_0) = \frac{1-\tan^2(\theta_0)}{1 + \tan^2(\theta_0)} = \frac{x_1^2 D_2 - x_2^2 D_1}{x_1^2 D_2 + x_2^2 D_1}
\end{equation}
(and similar for $\cos(2\Theta)$).  After simplifications, we finally
get a remarkably simple formula
\begin{equation}   \label{eq:T_halfline3}
\langle \T \rangle = \frac{x_1^2 - x_2^2}{2(D_2-D_1)}= \frac{\delta (2x_1 - \delta)}{2(D_2-D_1)} \,,
\end{equation}
which is valid for $x_1 > x_2$ and $D_1 < D_2$ (note that this result
could alternatively be derived by solving the Poisson equation
(\ref{eq:Tgeneral_def}) for the MFET).  This is precisely the MFPT for
a single particle with the initial position $x_2$ and diffusivity $D_2
- D_1$ to a fixed absorbing endpoint $x_1$ of an interval $(0,x_1)$
with reflections at $0$
\footnote{
Eq. \eqref{eq:T_halfline3} can be easily found
from the corresponding Green function for mixed boundary conditions, or
by realizing that the problem is equivalent to that on an interval of
doubled length $2x_1$ and two fully absorbing endpoints, whose
solution is well-known from the literature, see e.g. \cite{Redner}.}.
%footnote
Thus, the mean FET for the problem with a slowly diffusing target
(started at $x_1$) is the same as the mean FET for the problem with a
fixed target at $x_1$ if the diffusivity $D_2$ of the rapidly
diffusing particle is replaced with $D_2 - D_1$.  However, this
equivalence is only manifested at the level of the mean FET since
already the variance is different for these two problems, as we show
below.

The crucial difference between two particles here is that the second
particle (started from $x_2 \in (0,x_1)$) remains bound to a finite
interval between $0$ and the first particle (started from $x_1$),
whereas the latter is bound to a half-line between the second particle
and infinity.  When $D_1 > D_2$, the first particle diffuses faster
and can undertake very long excursions whose contributions make the
mean FET infinite, as in the case of a single particle on the
half-line with a fixed absorbing endpoint.  In other words, as the
slower second particle does not typically ``catch'' the first one
until after a very long time, its diffusion is not relevant.  In
contrast, when the second particle diffuses faster ($D_2 > D_1$), the
first particle cannot efficiently ``run away'' from it, and this
setting is similar to diffusion on a finite interval, for which the
mean FET is finite.

According to Eq. (\ref{eq:Tkmean}), the long-time decay
(\ref{eq:S_Redner}) implies that the higher-order moment $\langle \T^k
\rangle$ exists if $\Theta < \pi/(4k)$ and is given by
\begin{align}  \nonumber
\langle \T^k \rangle
& = 4k \sum\limits_{n=1}^\infty \frac{1-(-1)^n}{\pi n} \sin(\nu_n(\theta_0 + \Theta))  \\ \nonumber
& \times \int\limits_0^\infty dt \, t^{k-1} \, R_{\nu_n}\biggl(\frac{r_0}{\sqrt{D_1t}}\biggr)  \\
& = 2k \biggl(\frac{r_0^2}{D_1}\biggr)^k \sum\limits_{n=1}^\infty \frac{1-(-1)^n}{\pi n} \sin(\nu_n(\theta_0 + \Theta)) \, \varrho_k(\nu_n) ,
\end{align}
where
\begin{equation}
\varrho_k(\nu) = \frac{\sqrt{2\pi}}{2^{3k+1}}
\int\limits_0^\infty \frac{dz}{z^{k+1/2}} e^{-z} \biggl(I_{\frac{\nu+1}{2}}(z) + I_{\frac{\nu-1}{2}}(z)\biggr) .
\end{equation}
For integer-order moment $k$, one can use the following Laplace
transform \cite{Prudnikov4} % p. 313
\begin{align}  \nonumber
& \int\limits_0^\infty dt \, e^{-pt}\, t^{\mu} \, I_\nu(at) =
\frac{a^\nu \Gamma(\mu+\nu+1)}{2^\nu p^{\mu+\nu+1} \Gamma(\nu+1)} \\  
& \times ~ _2F_1\biggl(\frac{\mu+\nu+1}{2}, \frac{\mu+\nu+2}{2}; \nu+1; \frac{a^2}{p^2}\biggr)
\end{align}
(where $_2F_1(a,b;c;z)$ is the Gauss hypergeometric function) to obtain
\begin{equation}
\varrho_k(\nu) = (k-1)! \, \prod\limits_{j=1}^k \frac{1}{\nu^2 - (2j)^2} \,,
\end{equation}
which completes the computation of the $k$-th order moment:
\begin{align}  \nonumber
\langle \T^k \rangle
& = 2 k! (r_0^2/D_1)^k \sum\limits_{n=1}^\infty \frac{1-(-1)^n}{\pi n} \sin(\nu_n(\theta_0 + \Theta))  \\
& \times \prod\limits_{j=1}^k \frac{1}{\nu_n^2 - (2j)^2} \,.
\end{align}
Using again the summation identity (\ref{eq:summation}), we get
\begin{align}  \nonumber
\langle \T^2 \rangle & = - \frac{r_0^2}{6D_1} \langle \T \rangle + \frac{(r_0^2/D_1)^2}{96}
\biggl(\frac{\cos(4\theta_0)}{\cos(4\Theta)} - 1 \biggr),
\end{align}
whence
\begin{equation}
\langle \T^2 \rangle = \frac{(x_1^2 - x_2^2)(5x_1^2 D_2 + 5x_2^2 D_1 - x_1^2 D_1 - x_2^2 D_2)}{12(D_2-D_1)(D_1^2 + D_2^2 - 6D_1 D_2)} .
\end{equation}
The variance then is
\begin{align}   \nonumber
\sigma_{\T}^2 & = \frac{x_1^4 - x_2^4}{6(D_2-D_1)^2} \\  \label{eq:sigmaT}
& \times \frac{6(x_1^2 - x_2^2) D_1 D_2 + (x_1^2 + x_2^2)(D_2^2 - D_1^2)}{(x_1^2+x_2^2)(D_1^2 + D_2^2 - 6D_1 D_2)} \,.
\end{align}
Note that the condition $\Theta < \pi/8$ (with $k = 2$) is equivalent
to $D_1/D_2 < 3 - 2\sqrt{2} \approx 0.1716$, which is precisely one of
the roots of the quadratic polynomial in the denominator.  When
$D_1/D_2$ approaches this value, the variance diverges, whereas the
mean remains finite.

As said earlier, this expression differs from the variance for an
effective problem of a particle diffusing with diffusion coefficient
$D_2 - D_1$ to a fixed target at $x_1$:
\begin{equation}
\sigma_{\T,\,{\rm eff}}^2 = \frac{x_1^4 - x_2^4}{6(D_2-D_1)^2} \,.
\end{equation}
Indeed, Eq. (\ref{eq:sigmaT}) can be written as
\begin{equation*}
\sigma_{\T}^2 = \sigma_{\T,\,{\rm eff}}^2 \biggl(1 - 2\frac{x_1^2 (1 - 6D_2/D_1) + x_2^2}{(x_1^2 +x_2^2) \bigl(1 - 6 D_2/D_1 + (D_2/D_1)^2\bigr)}\biggr).
\end{equation*}

\section{Interval}
\label{sec:interval}

In this section, we consider the FET problem for two particles
diffusing on an interval $\Omega = (0,L)$, which is equivalent to
anisotropic diffusion on the square $(0,L) \times (0,L)$.  As this
domain is bounded, the governing diffusion operator, $D_1
\partial^2/\partial x_1^2 + D_2 \partial^2/\partial x_2^2$, has a
discrete spectrum, its eigenfunctions form a complete basis in the
space of square-integrable functions on $\Omega \times \Omega$, and
the survival probability admits a spectral expansion \cite{Gardiner}.
In particular, the survival probability decays exponentially
\begin{equation}
S(t|x_1,x_2) \propto \exp(-t/T)  \qquad (t\to \infty),
\end{equation}
with the decay time $T(D_1,D_2)$, determined by the smallest
eigenvalue of the diffusion operator.  This behavior is in sharp
contrast with the power-law decay (\ref{eq:S_Redner}) for diffusion on
a half-line.

We treat separately two cases: $D_1 = D_2$, for which an exact
solution and much more advanced analysis are possible
(Sec. \ref{sec:int_equal}), and $D_1 \ne D_2$, for which we focus on
the long-time limit (Sec. \ref{sec:int_different}).

\subsection{Equal diffusivities}
\label{sec:int_equal}

We search for the distribution of the first-encounter time for two
particles diffusing with equal diffusivities, $D_1 = D_2$, on an
interval $(0,L)$ with reflecting endpoints.  As discussed in
Sec. \ref{sec:equations}, the $N=2$ problem is equivalent to
two-dimensional diffusion on the square $(0,L)\times (0,L)$ with
reflecting edges and absorbing diagonal.  Our previous assumption $x_1
> x_2$ implies that the particle is actually restricted to the
isosceles right triangle $\Omega = \{ 0 < x_1 < L, 0 < x_2 < x_1\}$
with reflecting edges and absorbing hypotenuse.  For this domain, one
can construct the eigenfunctions of the Laplace operator by
antisymmetrizing the known eigenfunctions on the square:
\begin{align}  \nonumber
u_{n_1,n_2}(x_1,x_2) & = c_{n_1,n_2} \bigl[\cos(\pi n_1 x_1/L) \cos(\pi n_2 x_2/L) \\
& - \cos(\pi n_2 x_1/L) \cos(\pi n_1 x_2/L) \bigr] ,
\end{align}
with the indices $0 \leq n_1 < n_2$, and the normalization
coefficients are
\begin{equation}
\label{normcoeff}
c_{n_1 n_2} =\frac{2}{L} \frac{1}{ \sqrt{(1+\delta_{n_10})(1+\delta_{n_20})}} \,.
\end{equation}
As this set of eigenfunctions is complete (see Appendix
\ref{sec:completeness} and
\cite{Pinsky85,McCartin02,McCartin08,Grebenkov13}), the survival
probability can be written as a spectral decomposition:
\begin{equation}
\label{modedec}
S(t|x_1,x_2) = \sum\limits_{n_1=0}^\infty \sum\limits_{n_2>n_1}^\infty b_{n_1,n_2} \, u_{n_1,n_2}(x_1,x_2) \, e^{-D_1 t\lambda_{n_1,n_2}},
\end{equation}
where $\lambda_{n_1,n_2} = \pi^2 (n_1^2 + n_2^2)/L^2$ and
\begin{align}  \nonumber
b_{n_1,n_2} &= \int\limits_\Omega dx_1 \, dx_2\, u_{n_1,n_2}(x_1,x_2) \\
& = \frac{2L^2 c_{n_1,n_2}  (1 - (-1)^{n_1+n_2})}{\pi^2 (n_2^2 - n_1^2)}  \,.
\end{align}
This expression is a particular form of the general antisymmetrized
expression for $N$ vicious walkers provided in \cite{Bray04}.  
An alternative spectral representation of the survival probability was
derived in \cite{Tejedor11}:
\begin{align}  \nonumber
& S(t|x_1,x_2) = \frac{4}{\pi^2} \sum\limits_{n_1=0}^\infty \sum\limits_{n_2=0}^\infty \, e^{-D_1 t \lambda'_{n_1,n_2}} \\  \label{modedec2}
& \quad \times \frac{\sin\bigl(\frac{\pi (n_1+1/2)(x_1 - x_2)}{L}\bigr) \, \sin\bigl(\frac{\pi (n_2+1/2)(x_1 + x_2)}{L}\bigr)}{(n_1+1/2)(n_2+1/2)} \,,
\end{align}
where $\lambda'_{n_1,n_2} = 2\pi^2 [(n_1+1/2)^2 + (n_2+1/2)^2]/L^2$.
The decay time in the long-time limit is simply
\begin{equation}
T(D_1,D_1) = \frac{1}{D_1 \lambda_{0,1}} = \frac{L^2}{\pi^2 D_1} \,.
\end{equation}
In the long-time limit, the double sum in Eq. \eqref{modedec} is
essentially determined by a single decay mode associated with
$T(D_1,D_1)$, and one obtains
\begin{equation}
\label{survmode}
S(t|x_1,x_2)\approx \frac{8}{\pi^2}\left\{\cos( \pi x_2/L)-\cos(\pi x_1/L)\right\} e^{-D_1 \pi^2 t/L}.
\end{equation}
The particular case $x_1=L$ is of special interest, since it allows
one to compare the result more directly with the target problem
introduced in Sec. \ref{sec:comparison_sea}.  The average of
Eq. \eqref{survmode} over a uniform distribution of $x_2$ on the
interval $[0,L]$ yields
\begin{equation}
\label{avsurv}
S_{\rm uniform}(t)\simeq \frac{8}{\pi^2}e^{-D_1 \pi^2 t/L} \qquad (t\to \infty).
\end{equation}

From Eq. (\ref{eq:H_def}), the probability density of the FET is given
as
\begin{align} \nonumber
H(t|x_1,x_2) & = D_1 \sum\limits_{n_1=0}^\infty \sum\limits_{n_2=n_1+1}^\infty b_{n_1,n_2} \, u_{n_1,n_2}(x_1,x_2) \, \lambda_{n_1,n_2} \\
& \times e^{-D_1 t\lambda_{n_1,n_2}} .
\end{align}
The mean and higher-order moments of FET can be obtained from
Eq. (\ref{eq:Tkmean}):
\begin{align}  \label{Tk}
\langle \T^k \rangle
& = k! \sum\limits_{n_1=0}^\infty \sum\limits_{n_2>n_1}^\infty \frac{b_{n_1,n_2}}{(D_1 \lambda_{n_1,n_2})^k} \, u_{n_1,n_2}(x_1,x_2) .
\end{align}
In Appendix \ref{sec:MFET}, we employ the summation technique to
evaluate one of the double sum.  In particular, we derive the
following expression for the mean FET (for $x_1 \geq x_2$):
\begin{align} \nonumber
& \langle \T \rangle = \frac{(x_1-x_2)(2L^2 + 6x_2(L-x_1) - (x_1-x_2)^2)}{12D_1 L} \\   \label{eq:Tmean}
& + \frac{L^2}{D_1} \sum\limits_{n=1}^\infty \frac{\cos(\pi n \frac{x_2}{L}) v_n(\frac{x_1}{L}) 
- \cos(\pi n \frac{x_1}{L}) v_n(\frac{x_2}{L})}{(\pi n)^3}  \,,
\end{align}
where $v_n(x) = \frac{\cosh(\pi n(1-x)) - (-1)^n\cosh(\pi n
x)}{\sinh(\pi n)}$ (for $x_1 < x_2$, one needs just to exchange $x_1$
and $x_2$).  An alternative spectral representation of the mean FET,
based on Eq. (\ref{modedec2}), was derived in \cite{Tejedor11}:
\begin{align}  \nonumber
\langle \T \rangle & = \frac{(x_1-x_2)(L - (x_1-x_2)/2)}{2D_1} \\  \nonumber
& - \frac{L^2}{D_1} \sum\limits_{k=0}^\infty \frac{\sin\bigl(\frac{\pi (k+1/2)(x_1-x_2)}{L}\bigr)}{\pi^3 (k+1/2)^3} \\
& \times \frac{\sinh\bigl(\frac{\pi (k+1/2) (x_1+x_2)}{L}\bigr) + \sinh\bigl(\frac{\pi (k+1/2) (2L-x_1-x_2)}{L}\bigr)}{\sinh(\pi(2k+1))} \,.
\end{align}
Tejedor {\it et al.} discussed approximations of this exact relation,
in particular, when both particles are initially close to an endpoint
of the interval \cite{Tejedor11}.

\begin{figure}
\centering
\includegraphics[width=85mm]{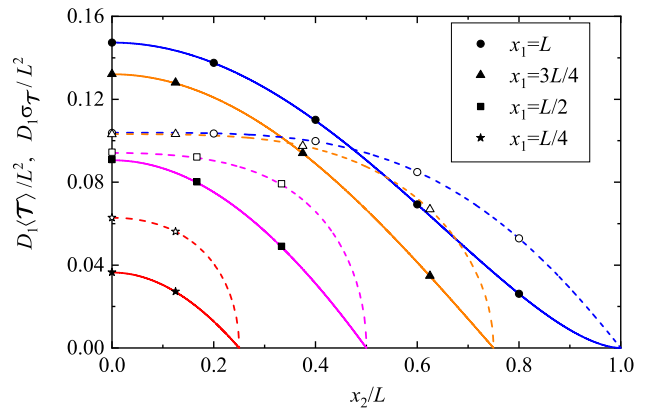} 
\caption{
Normalized mean FET (solid lines) and standard deviation (dashed
lines) versus the initial position of the left particle for several
initial positions of the right particle. The symbols show simulation
results with $10^6$ runs, obtained for $L=1000$ and particles with
diffusion coefficient $D_1=D_2=1/2$. }
\label{fig:t_sigmaD1=D2}
\end{figure}

\begin{figure}
\centering
\includegraphics[width=70mm]{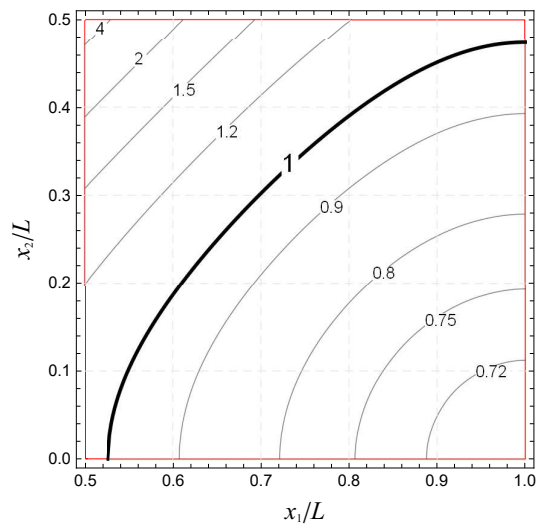}
\caption{
Contour plot of $\sigma_{\T}/\langle \T \rangle$ on the initial
position $x_1$--$x_2$ plane.  Contour lines corresponding to several
values of this ratio (which is shown as a label) are provided.  }
\label{fig:x1x2Regions}
\end{figure}

It is well known that the fluctuations in the values of $\T$ in many
first-passage time problems can be enormous
\cite{Redner,Levitz06,Mattos12,Grebenkov18}.  Equation \eqref{Tk}
allows us to compare $\langle \T \rangle$ and the standard deviation
$\sigma_{\T} = \sqrt{\langle \T^2\rangle - \langle \T \rangle^2}$ in
the present case of two particles in a interval
(Fig.~\ref{fig:t_sigmaD1=D2}).  We see that the standard deviation is
larger, and even much larger in some cases, than the mean FET when the
initial positions of the particles are close. For example, we see that
the standard deviation is always larger than the mean FET when both
particles start in the same half of the interval.  This is illustrated
with more detail in Fig.~\ref{fig:x1x2Regions} where a contour plot of
the ratio $\sigma_{\T}/\langle \T \rangle$ on the $x_1$--$x_2$ plane
is shown.

\subsubsection*{Comparison with the problem of a diffusing particle in a sea of diffusing traps}
\label{sec:comparison_sea}

By symmetry, the above problem with $x_1=L$ can be mapped to an
equivalent problem where the particle 1 starts at the middle of an
interval $(0,2L)$, and is surrounded by the particle 2 (which starts
at $x_2$ and has diffusivity $D_2=D_1$), and a fictitious mirror
particle 2' (which is originally located at $2L-x_2$ and follows
symmetrically the trajectory of the particle 2).  One could thus
wonder to what extent this problem is similar to other problems, in
which the distance to the nearest neighbors on a one-dimensional
setting determines the survival probability of the particle.

One such problem is the computation of the survival probability of a
diffusing point-like target with diffusivity $D_1$ in a sea of
identical noninteracting point traps, each of them diffusing with
diffusivity $D_2=D_1$. The traps are initially scattered at random
with a global density $\rho$ on the infinite real line.  In the limit
$t\to\infty$, it can be shown that the survival probability of the
target averaged over an ensemble of initial conditions is $S_{\rm
target}(t)\simeq e^{-4\rho (D_2 t/\pi)^{1/2}}$
\cite{Blythe2003,Yuste08,Borrego09}.  Taking the density value
$\rho=1/L$ and $D_2=D_1$ yields
\begin{equation}
\label{trappingsurv}
S_{\rm target}(t)\simeq  e^{-4 (D_1 t/\pi)^{1/2}/L} \qquad (t\to\infty),
\end{equation}
i.e, an asymptotic decay that is slower than the one prescribed by
Eq. \eqref{avsurv}.

In the target problem, there are large trap density fluctuations which
entail the formation of large gaps between the target and the closest
traps in certain statistical realizations. In contrast, the maximum
distance from particle 2 ({\it and} its mirror particle 2') to the
target at any time can never exceed the value $L$ in our expanded
system. This accelerates the decay of the survival probability with
respect to that observed in the target problem, despite the fact that
in the latter more than just two particles (actually, an infinite
number of them) are available to kill the target, since the traps can
overpass each other.

Finally, it is also worth noting that the decay observed in our
one-dimensional problem is similar to that observed in the
three-dimensional target problem with diffusing target of finite
extent \cite{PartII}; in both cases, the decay is exponential, since
the reflecting boundaries and the increased dimensionality
respectively facilitate the mixing of the reactants and accelerate the
decay of the survival probability.

\subsection{Unequal diffusivities}
\label{sec:int_different}

Here we investigate the mean FET $\langle \T \rangle$ and the decay
time $T(D_1,D_2)$ for the case $D_1\neq D_2$.  In particular, we will
consider the cases in which one of the diffusivities is much smaller
than the other, say $\epsilon^2=D_2/ D_1 \ll 1$.

\subsubsection{Mean FET}

\begin{figure}
\centering
\includegraphics[width=88mm]{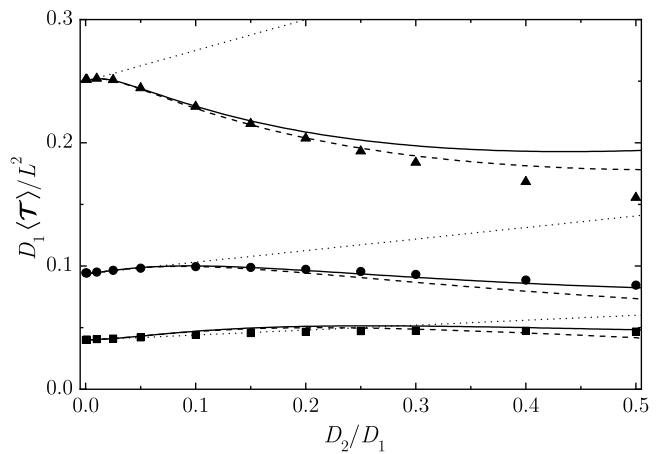}
\caption{
Scaled mean FET $\langle \T \rangle$ vs the ratio $D_2/D_1$ (with
$D_1=0.5$) for $\{x_1/L=0.75,x_2/L=0.25\}$ (triangles, $L=500$),
$\{x_1/L=0.75,x_2/L=0.5\}$ (circles, $L=500$) and
$\{x_1/L=0.9,x_2/L=0.7\}$ (squares, $L=1000$).  The symbols show
simulation results with $10^6$ runs.  The dotted line is the outer
solution~\eqref{uo}, the dashed line is the uniform
solution~\eqref{Tuniform} when one term is retained in the infinite
series, whereas the solid line is this same solution with ten terms
retained. }
\label{fig:TvsD2D1}
\end{figure}

In Ref.~\cite{Tzou2014}, Tzou {\it et al.} studied
Eq.~\eqref{eq:Tgeneral_def} and found the following asymptotic
expression for the mean FET for $\epsilon \ll 1$:
\begin{equation}
\label{Tuniform}
\frac{D_1\langle \T \rangle}{L^2} =u^o(\bar x_1,\bar x_2,\epsilon)+\epsilon V_1(\bar x_1,\bar x_2/\epsilon)+\epsilon^2 V_2(\bar x_1,\bar x_2/\epsilon),
\end{equation}
where
\begin{equation}
\label{uo}
u^0(\bar x_1,\bar x_2,\epsilon)=
\left(1+\epsilon^2\right)\left(\bar x_1-\bar x_2-\frac{\bar x_1^2}{2}+\frac{\bar x_2^2}{2}\right)
\end{equation}
is the outer solution, and
\begin{align}
\label{x}
V_1(x,\eta)&=-\sum_{n=0}^\infty \frac{2}{\alpha_n^2}e^{-\alpha_n \eta}\sin(\alpha_n x),\\
V_2(x,\eta)&=\int_0^\infty \frac{d\omega \, \hat F(\omega)}{\cosh(2\pi \omega)} \cosh\left[2\pi \omega(x-1)\right] \cos(2\pi \omega \eta),
\end{align}
with $\hat F(\omega)$ being the Fourier cosine transform of
\begin{equation}
\label{xf}
f(\eta)=\eta \sum_{n=0}^\infty \frac{2}{\alpha_n^2}e^{-\alpha_n \eta} \,,
\end{equation}
$\alpha_n=(n+1/2)\pi$, and $\bar x=x/L$.  In Fig.~\ref{fig:TvsD2D1} we
compare these expressions with simulation results. We see that when
the diffusivity of one of the particles is much smaller than the
diffusivity of the other, i.e., for $D_2\ll D_1$, Eq.~\eqref{uo} is a
simple and accurate expression for estimating the mean FET, $\langle
\T \rangle$, especially when the two particles start close to
each other.

In the limiting case of $D_2\to 0$, i.e., for $\epsilon\to 0$, one
gets
\begin{equation}
\label{x2}
\langle \T \rangle = \frac{2L (x_1- x_2) - x_1^2 + x_2^2}{2D_1} \,,
\end{equation}
which is just the mean FPT of a diffusive particle with diffusion
coefficient $D_1$ that starts at $x_1$ and is surrounded by an
absorbing frontier at $x_2$ and a reflecting barrier at $L$ (see
Eq.~\eqref{eq:T_halfline3} and the discussion below this equation).

Note that Tzou {\it et al.} in Ref.~\cite{Tzou2014} did not assess
the accuracy of their asymptotic formulas. They were mainly interested
by the question whether, in order to survive, it is better for one of
the particles to move randomly or remain immobile.

\subsubsection{Decay time}

\begin{figure}
\centering
\includegraphics[width=88mm]{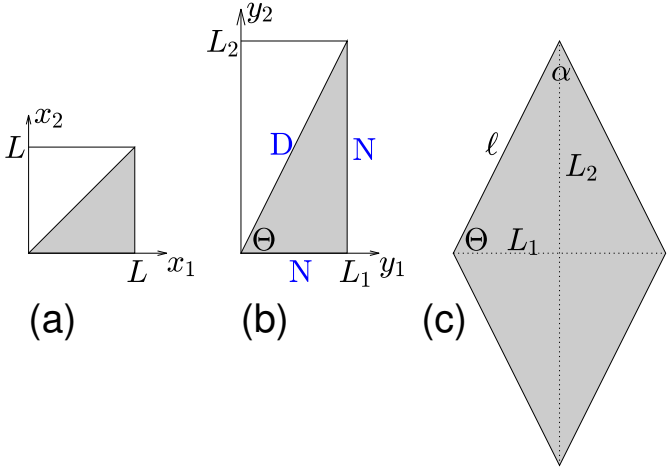}
\caption{
Schematic illustration of the first-encounter problem for two
particles with unequal diffusivities $D_2 < D_1$. {\bf (a)} The
original first-encounter problem is mapped onto anisotropic diffusion
in a square $(0,L)\times (0,L)$ with diffusivities $D_1$ and $D_2$
along coordinates $x_1$ and $x_2$.  The condition $x_1 > x_2$
restricts the starting point to the isosceles right triangle (gray
region).  {\bf (b)} In new variables $y_1 = x_1$ and $y_2 = x_2
\sqrt{D_1/D_2}$, this problem is equivalent to isotropic diffusion
in the right triangle with legs $(0,L_1)$ and $(0,L_2)$, where $L_1 =
L$ and $L_2 = L \sqrt{D_1/D_2}$.  The larger angle of the triangle is
$\Theta = \atan(L_2/L_1) = \atan(\sqrt{D_1/D_2})$.  Neumann (N)
boundary condition is imposed on both legs whereas Dirichlet (D)
boundary condition is imposed on the hypotenuse. {\bf (c)} Due to the
reflection symmetry of the ground eigenfunction of the Laplace
operator, the smallest Laplacian eigenvalue $\lambda_{\rm min}$ in the
above triangle can be determined from that in the rhombus with the
size $\ell = \sqrt{L_1^2 + L_2^2}$ and the acute angle $\alpha = \pi -
2\Theta$, where Dirichlet boundary condition is imposed on all edges.}
\label{fig:scheme}
% A_Felipe_triangle_fig;
\end{figure}

In the case $D_1 \neq D_2$, one can still stretch the original square
along one coordinate into a rectangle in order to get isotropic
diffusion (Fig. \ref{fig:scheme}).  In particular, the smallest
eigenvalue $\lambda_{\rm min}$ of the Laplace operator in the right
triangle with (reflecting) Neumann boundary conditions on legs
$(0,L_1)$ and $(0,L_2)$ (with $L_1 = L$ and $L_2 = L\sqrt{D_1/D_2}$
resulting from stretching) and (absorbing) Dirichlet boundary
condition on the hypotenuse determines the decay time $T$
\begin{equation}  \label{eq:T_lambda0}
T(D_1,D_2) = \frac{1}{D_1 \lambda_{\rm min}(D_1,D_2)} \,.
\end{equation}
Unfortunately, the eigenvalues and eigenfunctions of the Laplace
operator are not known for arbitrary right triangles.

However, the limiting case of one slowly diffusing particle, $D_2 \ll
D_1$, can be worked out. Note that, in this case, one deals with
strongly elongated triangles.  Due to Neumann boudary conditions on
the two legs, the original right triangle can be quadrupled by double
reflection along each leg, yielding a rhombus with absorbing boundary
condition on all four sides.  This operation does not change the
smallest eigenvalue.  As $D_2 \ll D_1$, the obtuse angle $2\Theta$ of
the rhombus is close to $\pi$, whereas the acute angle $\alpha =
2\Theta_0 = \pi - 2\Theta = 2\, \atan(\sqrt{D_2/D_1})$ is small.  In
\cite{Freitas07}, the following asymptotic behavior for the smallest
eigenvalue was derived:
\begin{equation}
\lambda_{\rm min} \simeq \frac{\pi^2}{\ell^2 \alpha^2} \biggl(1 - \frac{2^{2/3} a'_1}{\pi^{2/3}} \alpha^{2/3} + O(\alpha^{4/3})\biggr),
\end{equation}
where $\ell$ is the side length, $\alpha\ll 1$ is the angle, and $a'_1
\approx -1.0188$ is the first zero of the derivative of the Airy
function.  In our setting, $\ell^2 = L_1^2 + L_2^2 = L^2(1 + D_1/D_2)$
and $\alpha = 2\Theta_0$ so that
\begin{align}  \nonumber
\lambda_{\rm min} & \simeq \frac{\pi^2}{4L^2 (1 + D_2/D_1)} \frac{D_2/D_1}{\atan^2(\sqrt{D_2/D_1})} \\  \label{eq:lambda0_rect}
& \times \biggl(1 - \frac{2^{4/3} a'_1}{\pi^{2/3}}\,  \atan^{2/3}(\sqrt{D_2/D_1}) + \ldots\biggr) .
\end{align}
Expectedly, the smallest eigenvalue multiplied by $L^2$ is just a
function of $D_2/D_1$.  As a consequence, the decay time reads
\begin{align} \nonumber
T(D_1,D_2) & \simeq \frac{4L^2}{\pi^2 D_1} \, \frac{(D_1 + D_2)\atan^2(\sqrt{D_2/D_1})}{D_2}  \\  \label{eq:Trect}
& \times \biggl(1 + \frac{2^{4/3} a'_1}{\pi^{2/3}}\, \atan^{2/3}(\sqrt{D_2/D_1}) + \ldots\biggr) .
\end{align}
In the limit $D_2 \to 0$, the decay time approaches a constant,
$T(D_1,0^+) = 4L^2/(\pi^2 D_1)$.  We emphasize that this limit is
different from the one obtained in the case of a static target fixed
at $x_2$ and a particle diffusing on the interval $(x_2,L)$, for which
the decay time is $T(D_1,0) = 4(L-x_2)^2/(\pi^2 D_1)$. In other words,
the limit $D_2\to 0$ is singular, i.e.
\begin{equation}
\lim\limits_{D_2\to 0} T(D_1,D_2) = T(D_1,0^+) \ne T(D_1,0).
\end{equation}
In fact, even when $D_2$ is very small but strictly positive, the
memory of the starting position $x_2$ of the slow particle is lost in
the long-time limit, implying that $x_2$ does not influence the
timescale of the slowest decaying mode.  In this respect, the decay
time $T(D_1,D_2)$ is considerably different from the mean FET, which
depends on both starting point $x_1$ and $x_2$, see
Fig. \ref{fig:TvsD2D1}.

\begin{figure}
\centering
\includegraphics[width=88mm]{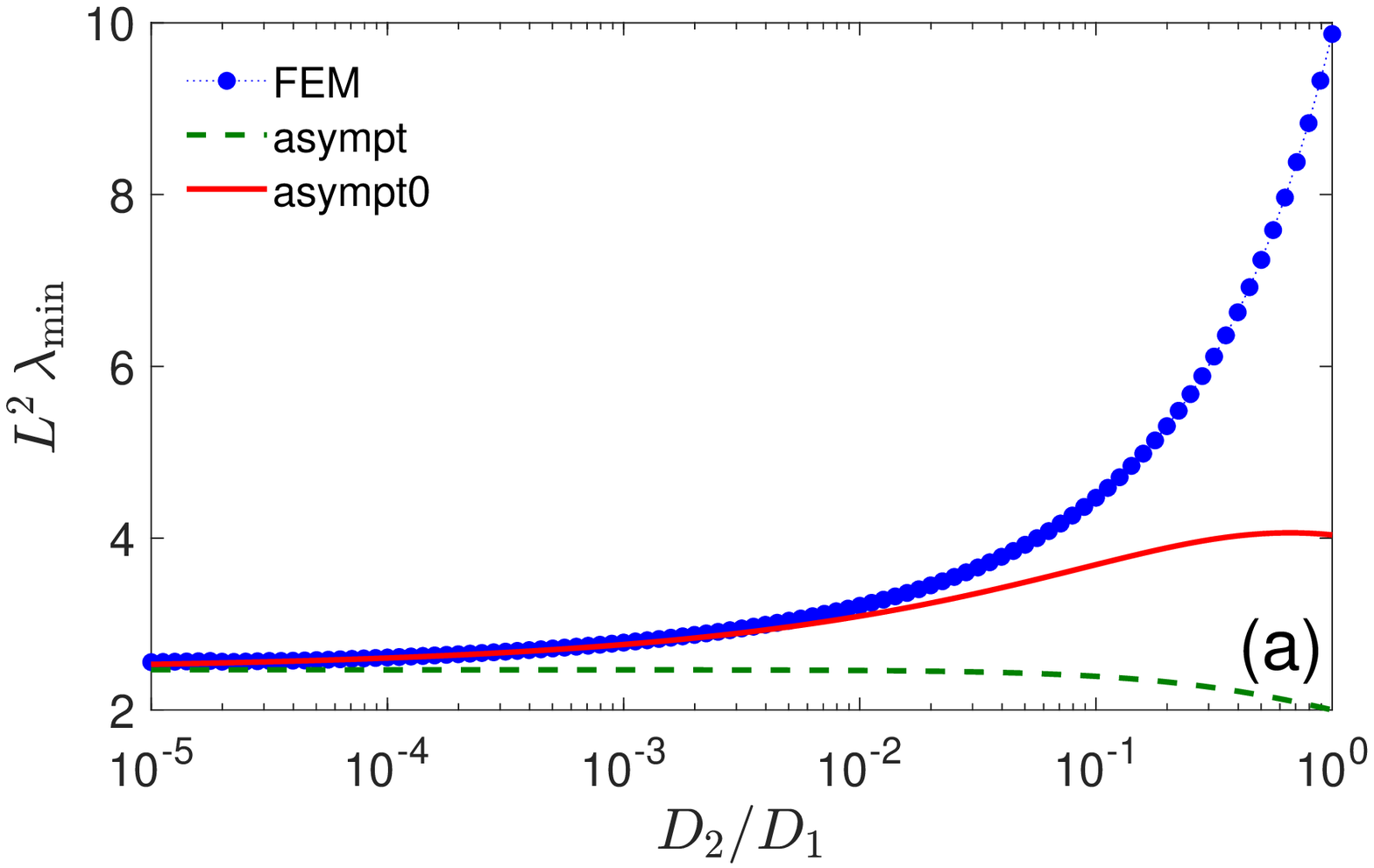}
\includegraphics[width=88mm]{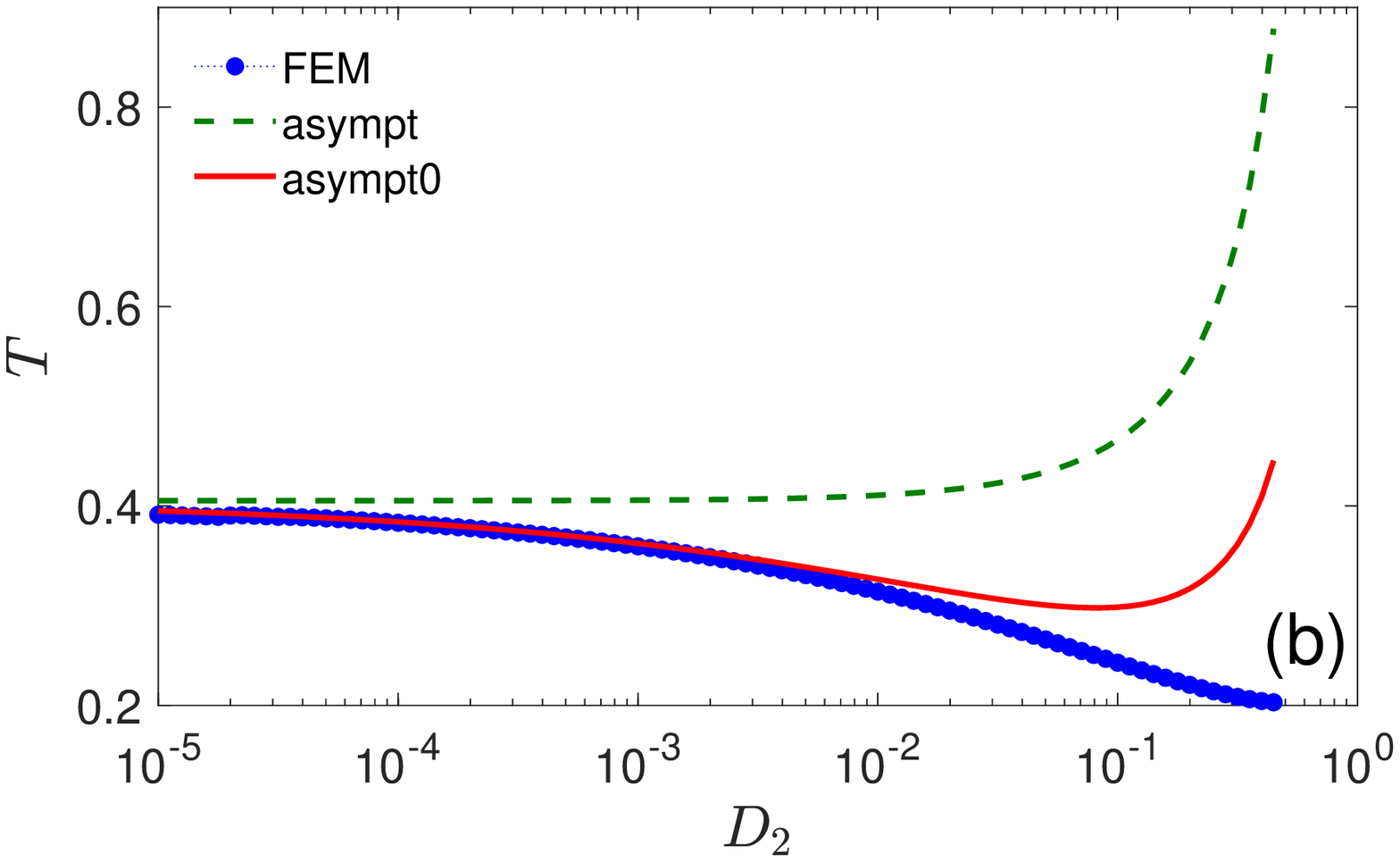}
\caption{
{\bf (a)} Smallest eigenvalue of the diffusion operator as a function
of $D_2/D_1$ for two particles diffusing on an interval $(0,L)$ with
diffusivities $D_1$ and $D_2$.  Filled circles show the eigenvalue
obtained numerically by a finite element method in Matlab PDEtool,
whereas solid and dashed lines represent the asymptotic formula
(\ref{eq:lambda0_rect}), with and without the subdominant term,
respectively.  {\bf (b)} Associated decay time $T$ vs. $D_2$ as
obtained from Eq. (\ref{eq:T_lambda0}) for $D_1 + D_2 = 1$ and $L =
1$.}
\label{fig:lambda0_rectangle}
% A_Felipe_FEM_rectangle_fignew;
% A_Felipe_FEM_rectangle_fignew2;
\end{figure}

Figure \ref{fig:lambda0_rectangle}(a) shows the behavior of the
smallest eigenvalue $\lambda_{\rm min}$ as a function of $D_2/D_1$.
At $D_2 = D_1$, we recover the square case considered in
Sec. \ref{sec:int_equal}, with $L^2 \lambda_{\rm min} = \pi^2$.  In
turn, as $D_2$ decreases, $L^2 \lambda_{\rm min}$ also decreases and
reaches the value $\pi^2/4$.  One can see that the asymptotic formula
(\ref{eq:lambda0_rect}) accurately captures the behavior of
$\lambda_{\rm min}$ for $D_2/D_1 \lesssim 0.01$.  It is worth noting
that the next-order correction term appearing in the second line of
(\ref{eq:lambda0_rect}) is necessary because the leading term alone
(dashed line) fails to reproduce the behavior.  Figure
\ref{fig:lambda0_rectangle}(b) further illustrates that
$T(D_1,D_2)$ is not a function of $D_1 + D_2$ alone (as in the
no-boundary case) but depends on both $D_1$ and $D_2$ in a more
intricate fashion.

\section{Conclusions}
\label{Sec:Conclusions}

In this paper, we investigated the impact of confinement onto the FET
distribution for two diffusing particles.  In spite of the practical
importance of this problem in chemical physics and related
disciplines, this problem has received little attention in comparison
with other first-passage problems, even for one-dimensional systems.
We focused on two settings: a half-line and an interval.  In the
half-line case, the survival probability $S(t|x_1,x_2)$ and the
related probability density $H(t|x_1,x_2)$ of the FET were already
formally known but had not been studied in detail, in particular, in
the short-time limit.  We thus carried out a thorough analysis of both
short- and long-time asymptotic behaviors, as well as the comparison
to the free case (without reflecting endpoint at the origin).  In
addition, we derived and discussed the behavior of the mean FET and
its variance.

The case of an interval was even less studied.  Both $S(t|x_1,x_2)$
and $H(t|x_1,x_2)$, as well as the moments of the FET, can be written
in terms of spectral expansions in the case of equal diffusivities.
We compared the mean FET and its standard deviation, in order to
quantify the role of fluctuations of the FET.  We also compared the
behavior for two particles with the problem of a diffusing particle in
a sea of diffusing traps, and found a faster decay of the survival
probability in the former case.  Finally, we investigated the case of
unequal diffusivities in the limit $D_2 \ll D_1$.  First, we checked
the quality of the asymptotic approximation for the mean FET reported
in \cite{Tzou2014}.  Second, we obtained another asymptotic relation
for the decay time characterizing the survival probability and the
probability density in the long-time limit.

As shown by our results, geometric confinement implies the onset of
additional time scales associated with the diffusion times of the
particles to the reflecting boundaries. Even in the case of a single
boundary, subtle effects emerge, e.g., the FET probability density may
be peaked at times earlier or later than in the no boundary case
depending on the parameter choice.  In the case when the particle 1
diffuses slowly, the mean FET is equal to the mean FPT for a single
particle with diffusivity $D_2-D_1$ moving between the origin and an
absorbing point at $x_1$, but the variances are different.  In the
presence of two boundaries (an interval), fluctuations of the FET can
also be important.  Moreover, we showed the mean FET and the decay
time are different and exhibit sophisticated dependences on both
diffusivities $D_1$ and $D_2$.  This observation breaks a common
intuitive thought, inspired by the no-boundary case, that only $D_1 +
D_2$ matters.  Finally, we illustrated that the limit $D_2 \to 0$ is
singular by deriving the asymptotic behavior of the decay time.  In
other words, the long-time behavior of the survival probability is
different for an immobile target ($D_2 = 0$) and for a very slowly
diffusing target ($D_2 \approx 0$).  This observation may question
common assumptions of static targets in biological systems, in which
everything is moving.

As we have seen, the additional scales introduced by boundaries result
in the onset of very rich behavior and drastic modifications with
respect to the free case. Even in simple settings, it is often not
possible to obtain exact analytic results, as exemplified by the
computation of the dominant decay mode in the interval problem.  From
a broader perspective, a variety of processes (fluorescence,
phosphorence and luminiscence quenching, reactions of solvated
electrons, proton transfer, radical recombination reactions,
enzyme-ligand interactions, etc.) have been shown to display reaction
rates that are often of the same order of magnitude as the predictions
of Smoluchowski's theory \cite{Smoluchowski1917,Rice}, but still
display important deviations.  Assessing the role of boundary effects
in some of these systems may help to better quantify these
discrepancies.

\begin{acknowledgments}
E.~A., F.~L.~V., and S.~B.~Y. acknowledge support by the Spanish
Agencia Estatal de Investigaci\'on Grant (partially financed by the
ERDF) No.~FIS2016-76359-P and by the Junta de Extremadura (Spain)
Grant (also partially financed by the ERDF) No.~GR18079. Additionally,
F.~L.~V. acknowledges financial support from the Junta de Extremadura
through Grant No. PD16010 (partially financed by ESF
funds). D.~S.~G. acknowledges a partial financial support from the
Alexander von Humboldt Foundation through a Bessel Research Award.
\end{acknowledgments}

\appendix
\section{Completeness of the eigenbasis for the isosceles right triangle}
\label{sec:completeness}

In this Appendix, we prove that the Laplacian eigenbasis used in
Sec. \ref{sec:int_equal} is complete.  Even so this result should be
known, we could not find its proof in the literature.

The starting point of the proof is the fact that the functions
\begin{eqnarray}
\phi_{n_1n_2}(x_1,x_2)&=& c_{n_1 n_2} \cos\left(\pi n_1 x_1/L\right) \cos\left(\pi n_2 x_2/L\right), \nonumber \\
&& \quad (n_1,n_2=0,1,2,\ldots),
\end{eqnarray}
with Eq. \eqref{normcoeff}, form a complete set of (orthonormal)
eigenfuctions of the Laplace operator on a square of side $L$ with
reflecting boundaries.  In other words, any square-integrable
function $f$ on the square can be decomposed onto this basis:
\begin{equation}
\label{fouSerie}
f(x_1,x_2)=\sum_{n_1=0}^\infty \sum_{n_2=0}^\infty b_{n_1 n_2} \phi_{n_1n_2}(x_1,x_2),
\end{equation}
where
\begin{equation}
b_{n_1 n_2}=\langle \phi_{n_1n_2}|f \rangle=\int_0^L\int_0^L \phi_{n_1n_2}(x_1,x_2) f(x_1,x_2) dx_1 dx_2.
\end{equation}
Let us now assume that $f$ has the symmetry $f(x_1,x_2)=-f(x_2,x_1)$.
This symmetry has the following implications on the values of the
Fourier coefficients $b_{n_1 n_2}$:

$\bullet$ $b_{nn}=0$.  In fact, by definition
\begin{equation}
\label{bnn}
b_{n  n }=\langle \phi_{n n }|f \rangle=\frac{2}{L}\int_0^L\int_0^L  F(x_1,x_2) dx_1 dx_2 ,
\end{equation}
where, for $n\neq 0$,
\begin{equation}
F(x_1,x_2) = \cos\left(\pi n  x_1/L\right)  \cos\left(\pi n x_2/L\right)  f(x_1,x_2) .
\end{equation}
But note that, due to the symmetry of $f$, $F(x_1,x_2)=-F(x_2,x_1)$.
This implies that the integral $\int_0^L\int_0^L F(x_1,x_2) dx_1 dx_2$
over the lower triangle $\Omega=\{0\le x_1\le L, 0\le x_2\le x_1\}$ is
equal (but with opposite sign) to the integral over the upper triangle
$\bar \Omega=\{0\le x_1\le L, 0\le x_1\le x_2\}$. Therefore,
$b_{nn}=0$.  The proof for $n=0$ is straightforward.

$\bullet$ $b_{n_1 n_2}=-b_{n_2 n_1}$.  In fact, by definition
\begin{align}
& b_{n_2 n_1}=\nonumber \\
& \frac{2}{L} \int_0^L\int_0^L \cos\left(\pi n_2 x_1/L\right)  \cos\left(\pi n_1 x_2/L\right) f(x_1,x_2) dx_1 dx_2
\end{align}
or, using the property  $f(x_1,x_2)=-f(x_2,x_1)$,
\begin{align}
& b_{n_2 n_1}= \nonumber \\
&-\frac{2}{L} \int_0^L\int_0^L   \cos\left(\pi n_1 x_2/L\right) \cos\left(\pi n_2 x_1/L\right) f(x_2,x_1) dx_2 dx_1 \nonumber \\
&=-b_{n_1 n_2} .
\end{align}
This is also true if $n_1=0$ or $n_2=0$.

Using the results $b_{n_1 n_2}=-b_{n_2 n_1}$ and $b_{nn}=0$, one finds
that any function $f(x_1,x_2)$ with the property
$f(x_1,x_2)=-f(x_2,x_1)$ can be uniquely represented in terms of the
eigenfunctions $u_{n_1,n_2}(x_1,x_2)=\phi_{n_1 n_2}(x_1,x_2)
-\phi_{n_2 n_1}(x_1,x_2)$ with $0\le n_1<n_2$. Note, however, that the
property $f(x_1,x_2)=-f(x_2,x_1)$ does not imply any restriction on
the value of $f$ on the lower triangle $\Omega$. Thus, any
square-integrable function $f(x_1,x_2)$ defined on $\Omega$ can be
uniquely represented in terms of the eigenfunctions
$u_{n_1,n_2}(x_1,x_2)=\phi_{n_1 n_2}(x_1,x_2) -\phi_{n_2
n_1}(x_1,x_2)$ with $0\le n_1<n_2$.  In other words,
$u_{n_1,n_2}(x_1,x_2)$ with $0\le n_1<n_2$ form a complete set of
Laplacian eigenfunctions on $\Omega$.

\section{MFET for the interval}
\label{sec:MFET}

In this Appendix, we present a lengthy and technical derivation of the
mean FET of two particles diffusing with equal diffusivities on the
interval $(0,L)$ with reflecting endpoints.  We start by rewriting
Eq. (\ref{Tk}) explicitly as
\begin{align*}
& \langle \T^k \rangle
= C_k \hspace*{-1mm} \sum\limits_{n_1=0}^\infty \sum\limits_{n_2> n_1}  \hspace*{-1.5mm}
\frac{1-(-1)^{n_1+n_2}}{(1+\delta_{n_1,0})(\alpha_{n_2}^2-\alpha_{n_1}^2)(\alpha_{n_1}^2 + \alpha_{n_2}^2)^k} \\
& \times \biggl(\cos(\alpha_{n_1} x_1) \cos(\alpha_{n_2} x_2) - \cos(\alpha_{n_1} x_2) \cos(\alpha_{n_2} x_1)\biggr),
\end{align*}
where $\alpha_n = \pi n$, $C_k = 8 (k!) (L^2/D_1)^k$, and we rescaled
$x_1$ and $x_2$ by $L$ for shorter notations.  Note that here we
assumed that $x_1 \geq x_2$.  As this expression is antisymmetric with
respect to exchange $x_1 \leftrightarrow x_2$, one would need to
change the sign for $x_1 < x_2$.

We first separate the term with $n_1 = 0$, for which we get
\begin{align}
S_k^{(0)} & = \frac{C_k}{2} \sum\limits_{n_2=1}^\infty \frac{1-(-1)^{n_2}}{\alpha_{n_2}^{2(k+1)}}
\biggl(\cos(\alpha_{n_2}x_2) - \cos(\alpha_{n_2}x_1)\biggr).
\end{align}
We use the summation identities \cite{Grebenkov20}
\begin{subequations} \label{eq:identities}
\begin{align}  \label{eq:identities1}
\sum\limits_{n=1}^\infty \frac{\cos(\alpha_n x) }{s + \alpha_{n}^2} 
& = \frac{\cosh(\sqrt{s}(1-x))}{2\sqrt{s} \sinh(\sqrt{s})} - \frac{1}{2s} \,, \\  \label{eq:identities2}
\sum\limits_{n=1}^\infty \frac{\cos(\alpha_n x) (-1)^n}{s + \alpha_{n}^2} 
& = \frac{\cosh(\sqrt{s} x)}{2\sqrt{s} \sinh(\sqrt{s})} - \frac{1}{2s}  \,,
\end{align}
\end{subequations}
to compute
\begin{align}  \label{eq:Fsx}
F(s,x) & = \sum\limits_{n=1}^\infty \frac{(1-(-1)^{n}) \cos(\alpha_n x) }{s + \alpha_{n}^2} \\  \nonumber
& = \frac{\cosh(\sqrt{s}(1-x)) - \cosh(\sqrt{s} x)}{2\sqrt{s} \sinh(\sqrt{s})} \,.
\end{align}
Evaluating the $k$-th derivative of this identity at $s = 0$, one gets
\begin{equation}
S_k^{(0)} = \frac{C_k \, (-1)^k}{2(k!)} \lim\limits_{s\to 0} \frac{\partial^k [F(s,x_2)-F(s,x_1)]}{\partial s^k}  \,.
\end{equation}

Now we switch to the evaluation of the double sum with $n_1 > 0$ and
$n_2 > n_1$:
\begin{align*}
& S_k^{(1)} = \frac{C_k}{2}\sum\limits_{n_1=1}^\infty \sum\limits_{n_2 \ne n_1} \frac{1-(-1)^{n_1+n_2}}{\alpha_{n_2}^2-\alpha_{n_1}^2}\,
\frac{1}{(\alpha_{n_1}^2 + \alpha_{n_2}^2)^k} \\
& \times \biggl(\cos(\alpha_{n_1} x_1) \cos(\alpha_{n_2} x_2) - \cos(\alpha_{n_1} x_2) \cos(\alpha_{n_2} x_1)\biggr),
\end{align*}
where we employed the symmetry of the summand expression with respect
to exchange $n_1 \leftrightarrow n_2$ to symmetrize the second sum.
Our goal is to evaluate exactly the second sum over $n_2$:
\begin{align*}
& W_{k,n_1} = \sum\limits_{n_2 \ne n_1} \frac{1-(-1)^{n_1+n_2}}{\alpha_{n_2}^2-\alpha_{n_1}^2}\,
\frac{1}{(\alpha_{n_1}^2 + \alpha_{n_2}^2)^k} \\
& \times \biggl(\cos(\alpha_{n_1} x_1) \cos(\alpha_{n_2} x_2) - \cos(\alpha_{n_1} x_2) \cos(\alpha_{n_2} x_1)\biggr),
\end{align*}
so that
\begin{equation}
S_k^{(1)} = \frac{C_k}{2} \sum\limits_{n_1=1}^\infty W_{k,n_1} \,.
\end{equation}
For this purpose, we evaluate the following sum:
\begin{align*}
U_k(s,x) & = \sum\limits_{n=1}^\infty \frac{\cos(\alpha_n x) (-1)^n}{(\alpha_{n}^2-s)(\alpha_n^2 + s)^k}  \,.
\end{align*}
Using the identity
\begin{equation}
\frac{1}{(\alpha_n^2-s)(\alpha_n^2+s)^k} = \frac{1}{(2s)^k(\alpha_n^2-s)} - \sum\limits_{j=1}^{k} \frac{(2s)^{j-k-1}}{(\alpha_n^2+s)^{j}}
\end{equation}
we can evaluate this sum with the help of Eq. (\ref{eq:identities2}):
\begin{align*}
U_k(s,x) & = \frac{1}{(2s)^k} \biggl(-\frac{\cos(\sqrt{s} x)}{2\sqrt{s}\sin(\sqrt{s})} + \frac{1}{2s}\biggr) \\
& - \sum\limits_{j=0}^{k-1} (2s)^{j-k} \frac{(-1)^j}{j!} \frac{\partial^j}{\partial s^j} 
\biggl(\frac{\cosh(\sqrt{s} x)}{2\sqrt{s}\sinh(\sqrt{s})} - \frac{1}{2s}\biggr) \\
& = \frac{1}{2s^{k+1}} - \frac{1}{(2s)^k} \frac{\cos(\sqrt{s} x)}{2\sqrt{s}\sin(\sqrt{s})}  \\
& - \sum\limits_{j=0}^{k-1} (2s)^{j-k} \frac{(-1)^j}{j!} \frac{\partial^j}{\partial s^j} 
\frac{\cosh(\sqrt{s} x)}{2\sqrt{s}\sinh(\sqrt{s})} \,.\\
\end{align*}

Now we can come back to the sum $W_{n_1}$, which can be split into 4
terms:
\begin{align*}
& W_{k,n_1} = \cos(\alpha_{n_1} x_1) \bigl( V_{k,n_1}^{(1)}(x_2) - (-1)^{n_1} V_{k,n_1}^{(2)}(x_2)\bigr) \\
& - \cos(\alpha_{n_1} x_2) \bigl(V_{k,n_1}^{(1)}(x_1) - (-1)^{n_1} V_{k,n_1}^{(2)}(x_1)\bigr) ,
\end{align*}
where
\begin{align}
V_{k,n_1}^{(1)}(x) & = \sum\limits_{n_2 \ne n_1} \frac{\cos(\alpha_{n_2} x)}{(\alpha_{n_2}^2-\alpha_{n_1}^2)(\alpha_{n_1}^2 + \alpha_{n_2}^2)^k} \,, \\
V_{k,n_1}^{(2)}(x) & = \sum\limits_{n_2 \ne n_1} \frac{\cos(\alpha_{n_2} x) 
(-1)^{n_2}}{(\alpha_{n_2}^2-\alpha_{n_1}^2)(\alpha_{n_1}^2 + \alpha_{n_2}^2)^k} \,.
\end{align}
These sums can be evaluated by using $U_k(s)$.  In fact, replacing
$\alpha_{n_1}^2$ by $s$ in the above expressions, one can first
evaluate these sums for $s \ne \alpha_{n_1}^2$ by adding and
subtracting the term $n_2 = n_1$, and then take the limit $s\to
\alpha_{n_1}^2$:
\begin{align*}
V_{k,n_1}^{(1)}(x) & = \lim\limits_{s\to \alpha_{n_1}^2} \biggl(U_k(s,1-x) 
- \frac{\cos(\alpha_{n_1}x)}{(\alpha_{n_1}^2-s)(s+\alpha_{n_1}^2)^k} \biggr), \\
V_{k,n_1}^{(2)}(x) & = \lim\limits_{s\to \alpha_{n_1}^2} \biggl(U_k(s,x) 
- \frac{\cos(\alpha_{n_1}x) (-1)^{n_1}}{(\alpha_{n_1}^2-s)(s+\alpha_{n_1}^2)^k} \biggr) .
\end{align*}
The subtracted term removes the singularity in $U_k(s,1-x)$ and
$U_k(s,x)$ as $s\to \alpha_{n_1}^2$.
This completes our formal evaluation of the moment $\langle
\T^k\rangle$, which is just the sum of $S_k^{(0)}$ and $S_k^{(1)}$
given above.

Let us apply this general evaluation to get the mean FET $\langle \T \rangle$.
For $k = 1$, we have $C_1 = 8L^2/D_1$ and
\begin{equation}
S_1^{(0)} = \frac{4}{D_1} \left(\frac{x_2^3-x_1^3}{12L} - \frac{x_2^2 - x_1^2}{8}\right).
\end{equation}
To evaluate the contribution $S_1^{(1)}$, we first find
\begin{equation}
U_1(s,x) = \frac{1}{4s} \biggl(\frac{2}{s} - \frac{\cos(\sqrt{s}x)}{\sqrt{s}\sin(\sqrt{s})} 
- \frac{\cosh(\sqrt{s}x)}{\sqrt{s}\sinh(\sqrt{s})} \biggr) .
\end{equation}
Then we compute the limit
\begin{align*}
V_{1,n_1}^{(2)}(x) & = \frac{1}{4\alpha_{n_1}^4} \biggl(2 - \frac{\alpha_{n_1}\cosh(\alpha_{n_1}x)}{\sinh \alpha_{n_1}} \\
& + (-1)^{n_1} \biggl(\frac32 \cos(\alpha_{n_1}x) + x \alpha_{n_1} \sin(\alpha_{n_1}x)\biggr)\biggr),
\end{align*}
and $V_{1,n_1}^{(1)}(x) = V_{1,n_1}^{(2)}(1-x)$.  Combining these
results, we get
\begin{equation}
W_{1,n} = \frac{\cos(\alpha_n x_1) w_n(x_2) - \cos(\alpha_n x_2) w_n(x_1)}{4\alpha_n^4}  \,,
\end{equation}
where
\begin{align*}
w_n(x) & = 2(1-(-1)^n) - \alpha_n \sin(\alpha_n x) \\
& - \frac{\alpha_n}{\sinh \alpha_n} \bigl(\cosh(\alpha_n(1-x)) - (-1)^n \cosh(\alpha_n x)\bigr) .
\end{align*}
As a consequence, the above expression allows one to split $S_1^{(1)}$
into three contributions:
\begin{equation*}
S_1^{(1)} = S_1^{(1,1)} + S_1^{(1,2)} + S_1^{(1,3)} ,
\end{equation*}
where
\begin{align*}
S_{1}^{(1,1)} & = \frac{C_1}{2} \sum\limits_{n=1}^\infty \frac{(1-(-1)^n)(\cos(\alpha_n x_1) - \cos(\alpha_n x_2))}{2\alpha_n^4} \,, \\
S_{1}^{(1,2)} & = \frac{C_1}{2} \sum\limits_{n=1}^\infty \frac{\sin(\alpha_n (x_1-x_2))}{4\alpha_n^3}  \,, \\
S_{1}^{(1,3)} & = \frac{C_1}{2} \sum\limits_{n=1}^\infty \frac{\cos(\alpha_n x_2) v_n(x_1) - \cos(\alpha_n x_1) v_n(x_2)}{4\alpha_n^3} \,,
\end{align*}
with
\begin{equation}  \label{eq:vn}
v_n(x) = \frac{\cosh(\alpha_n(1-x)) - (-1)^n\cosh(\alpha_n x)}{\sinh \alpha_n} \,.
\end{equation}

Note that $S_1^{(1,1)} = -S_1^{(0)}/2$.  The second sum can be easily
computed by taking the derivative of Eq. (\ref{eq:identities1}) with
respect to $x$ and $s$ and evaluating the limit $s\to 0$:
\begin{equation}
\sum\limits_{n=1}^\infty \frac{\sin(\alpha_n x)}{\alpha_n^3} = \frac{x(1-x)(2-x)}{12} \,,
\end{equation}
from which
\begin{equation}
S_1^{(1,2)} = \frac{(x_1-x_2)(L-x_1+x_2)(2L-x_1+x_2)}{12D_1 L}  \,.
\end{equation}
In summary, we conclude for $x_1 \geq x_2$ that
\begin{align*}
& \langle \T \rangle = \frac{(x_1-x_2)(2L^2 + 6x_2(L-x_1) - (x_1-x_2)^2)}{12D_1L} + \\  \nonumber
& \frac{L^2}{D_1} \sum\limits_{n=1}^\infty \frac{\cos(\alpha_n x_2/L) v_n(x_1/L) - \cos(\alpha_n x_1/L) v_n(x_2/L)}{\alpha_n^3}  \,,
\end{align*}
with $v_n(x)$ given by Eq. (\ref{eq:vn}), and $\alpha_n = \pi n$.  For
$x_1 < x_2$, one needs just to exchange $x_1$ and $x_2$.

\end{document}